\documentclass[11pt,superscriptaddress,longbibliography,aps,showpacs,notitlepage,nofootinbib]{revtex4-1}
\usepackage{amssymb,amsmath}
\usepackage{bm}
\usepackage[pdftex]{graphicx}
\usepackage{subfig}
\usepackage{color}

\usepackage{blindtext}
\usepackage{graphicx}
\usepackage{natbib}
\usepackage{siunitx}
\usepackage{textcomp}
\usepackage{gensymb}
\usepackage{multirow}
\usepackage[dvipsnames]{xcolor}
\usepackage{hyperref}
\hypersetup{colorlinks=true,linkcolor=blue,filecolor=magenta,urlcolor=cyan}
\usepackage[nameinlink, noabbrev]{cleveref}
\usepackage{soul}

\newcommand{\beq}{\begin{equation}}
\newcommand{\eeq}{\end{equation}}

\begin{document}
	\title{
Time-Scales for Nonlinear Processes in Preheating after Multifield Inflation with Nonminimal Couplings
}

\newcommand{\Kenyon}{Department of Physics, Kenyon College, Gambier, Ohio 43022, USA}
\newcommand{\UIUC}{Department of Physics, University of Illinois at Urbana-Champaign, Urbana, Illinois 61801,
U.S.A.}
\newcommand{\Desy}{DESY, Notkestra{\ss}e 85, D-22607 Hamburg, Germany}
\newcommand{\Nikhef}{Nikhef, Science Park 105, 1098XG Amsterdam, The Netherlands}
\newcommand{\Lorentz}{Lorentz Institute for Theoretical Physics, Leiden University, 2333CA Leiden, The Netherlands}
\newcommand{\MIT}{Department of Physics, Massachusetts Institute of Technology, Cambridge, Massachusetts 02139, USA}
\newcommand{\Case}{CERCA/ISO, Department of Physics, Case Western Reserve University, 10900 Euclid Avenue, Cleveland, OH 44106}

\author{Jorinde~van~de~Vis}
\email{jorinde.van.de.vis@desy.de}
\affiliation{\Desy}
\author{Rachel~Nguyen}
\email{nguyenr@kenyon.edu}
\affiliation{\Kenyon}
\affiliation{\UIUC}
\author{Evangelos~I.~Sfakianakis}
\email{evans@nikhef.nl}
\affiliation{\Nikhef} 
\affiliation{\Lorentz}
\author{John~T.~Giblin,~Jr.}
\email{giblinj@kenyon.edu}
\affiliation{\Kenyon}
\affiliation{\Case}
\author{David~I.~Kaiser}
\email{dikaiser@mit.edu}
\affiliation{\MIT}

\begin{abstract}
{
We have conducted extensive lattice simulations to study the post-inflation dynamics of multifield models involving nonminimal couplings. We explore the parameter dependence of preheating in these models and describe the various time-scales that control such nonlinear processes as energy transfer, re-scattering, and the approach to radiation-domination and thermalization. In the limit of large nonminimal couplings ($\xi_I \sim 100$), we find that efficient transfer of energy from the inflaton condensate to radiative degrees of freedom, emergence of a radiation-dominated equation of state, and the onset of thermalization each consistently occur within $N_{\rm reh} \lesssim 3$ $e$-folds after the end of inflation, largely independent of the values of the other couplings in the models. The exception is the case of negative ellipticity, in which there is a misalignment between the dominant direction in field-space along which the system evolves and the larger of the nonminimal couplings $\xi_I$. In those cases, the field-space-driven parametric resonance is effectively shut off. More generally, the competition between the scalar fields' potential and the field-space manifold structure can yield interesting phenomena such as two-stage resonances. Across many regions of parameter space, we find efficient re-scattering between the distinct fields, leading to a partial memory loss of the shape of the initial fluctuation spectrum. Despite the explosive particle production, which can lead to a quick depletion of the background energy density, the nonlinear processes do not induce any super-horizon correlations after the end of inflation in these models, which keeps predictions for CMB observables unaffected by the late-time amplification of isocurvature fluctuations. Hence the excellent agreement between primordial observables and recent observations is preserved for this class of models, even when we consider post-inflation dynamics.
}
\end{abstract}

\date{\today}

\maketitle

\section{Introduction}

Inflation is the leading framework for studying the very early universe, and predictions from several inflationary models are consistent with high-precision cosmological observations, including the {\it Planck} measurements of the cosmic microwave background radiation (CMB) and large-scale structure surveys \cite{Guth:2013sya,Martin:2015dha,Chowdhury:2019otk}. However, calculations of predictions for observables from inflationary models are subject to uncertainty arising from the limited amount of information we have regarding the post-inflation reheating era. Reheating is the period during which the energy density that had been locked in the inflaton condensate is transferred to radiation modes that (will eventually) become the Standard Model and Dark Matter sectors. The expansion of the universe thus transitions from a quasi-de Sitter phase during inflaton to a radiation-dominated phase at the end of reheating. The duration of reheating and corresponding equation of state affect the relation between the times of horizon exit and re-entry for inflationary fluctuations \cite{Martin:2010kz,Mielczarek:2010ag,Martin:2014nya,Munoz:2014eqa,Creminelli:2014oaa,Creminelli:2014fca,Dai:2014jja,Cai:2015soa,Cook:2015vqa,Eshaghi:2016kne,Ueno:2016dim,Figueroa:2018twl}. Reheating is thus a crucial phase to understand, in order to connect the primordial phase of inflation with the conditions required for the onset of standard Big Bang evolution, as well as to refine predictions for observables from inflationary models. Reheating can proceed either through perturbative decays of the inflaton into lighter particles (perturbative reheating), or through nonperturbative processes such as parametric resonance (preheating), which yield an exponential enhancement of radiation modes. The latter process --- when efficient enough --- can lead to very fast reheating and the emergence of nonlinear effects, such as oscillon formation \cite{Amin:2010xe,Amin:2010dc,Gleiser:2011xj, Amin:2011hj,Lozanov:2017hjm,Hong:2017ooe,Fukunaga:2019unq}. (For reviews of reheating, see Refs.~\cite{Bassett:2005xm,Frolov:2010sz,Allahverdi:2010xz,Amin:2014eta,Lozanov:2019jxc}.)

In this work we study a family of models that incorporates realistic features from high-energy physics, namely multiple interacting scalar fields, each coupled nonminimally to the space-time Ricci scalar. Nonminimal couplings of the form $\xi \phi^2 R$, where $\xi$ is a dimensionless coupling, $\phi$ is a scalar field, and $R$ is the spacetime Ricci scalar, necessarily arise as renormalization counterterms for self-interacting scalar fields in curved spacetime \cite{Callan:1970ze,Bunch:1980br,Bunch:1980bs,Birrell:1982ix,Odintsov:1990mt,Buchbinder:1992rb,Parker:2009uva,Markkanen:2013nwa}. Furthermore, the values of nonminimal couplings typically increase with energy scale under renormalization-group flow, with no UV fixed points \cite{Odintsov:1990mt,Buchbinder:1992rb}.  We thus expect multiple scalar fields to be present during inflation, each with a potentially large nonminimal coupling. Higgs inflation~\cite{Bezrukov:2007ep,Greenwood:2012aj,Sfakianakis:2018lzf,Ema:2020zvg} is a well-known example from this class of models. 

This class of models yields an excellent fit between predictions and CMB observables, when such predictions are calculated based only on the dynamics {\it during} inflation \cite{Kaiser:2013sna}. Yet it is critical to understand the time-scales over which distinct processes during the post-inflation reheating phase unfold, such as the onset of a radiation-dominated equation of state and the emergence of thermal equilibrium at an appropriately high temperature. In this family of models in particular, one must also track the evolution of the single-field attractor \cite{Kaiser:2012ak,Greenwood:2012aj,Kaiser:2013sna,Schutz:2013fua} beyond the end of inflation in order to understand whether any post-inflation effects could significantly affect the predictions for primordial observables \cite{Nguyen:2019kbm}.

As a considerable amount of energy is transferred from the inflaton to radiation modes, the most significant physical processes during (p)reheating are essentially {\it nonlinear}. 
In order to capture non-linear processes during preheating, we conduct large-scale lattice simulations. We consider two-field models, with $\phi^I = \{ \phi, \chi \}$, and couplings $(\xi_\phi \phi^2 + \xi_\chi \chi^2 ) R$ in the Lagrangian. We build upon Ref.~\cite{Nguyen:2019kbm} to identify several distinct nonlinear effects, characteristic of the preheating dynamics, which can unfold on different time-scales. To explore the nonlinear dynamics of this family of models, we focus on several distinct processes and consider how the associated time-scales change as one varies the relevant parameters. We first consider a fairly ``generic" set of ratios among the couplings, and study the dynamics of this ``benchmark" case across three distinct regimes of $\xi_\phi \sim 1, 10, 100$.  We then consider how this behavior shifts for a model that has symmetric couplings as well as for models in which we vary the ellipticity, $\varepsilon \equiv (\xi_\phi - \xi_\chi) / \xi_\phi$. 

 Across these many regions of parameter space, we find a dominant trend that increasing the non-minimal couplings $\xi_I$ increases the efficiency of preheating, thereby shortening the time required for energy to be transferred from the inflaton condensate into a (quasi-) thermal bath of produced particles. Throughout parameter space for this family of models, we consistently find the emergence of a radiation-dominated equation of state as well as the onset of thermalization to occur within $N_{\rm reh} \sim {\cal O} (1)$ $e$-folds after the end of inflation. We further find that the single-field attractor that had been identified in previous work \cite{Kaiser:2012ak,Greenwood:2012aj,Kaiser:2013sna,Schutz:2013fua} persists through the end of the preheating phase in most cases, thereby protecting the close match between predictions from this family of models and the latest high-precision measurements of CMB observables. Even in the case in which the excitation of the second field is strong enough to break the single-field attractor, this occurs deep into preheating and only after nonlinear effects have rendered the notion of a background trajectory invalid. Furthermore, the average field trajectory does not exhibit super-horizon correlations, and hence cannot affect CMB observables.


In Section~\ref{sec:modeletc} we define the family of models and briefly discuss the behavior of fluctuations, which depend on both the curvature of the field-space manifold and on the topography of the potential. In Section~\ref{sec:benchmark} we examine a characteristic ``benchmark'' case and define the various time-scales that are relevant for understanding the non-linear dynamics during preheating. In Section~\ref{sec:parameters} we discuss how the efficiency of preheating and the dynamical time-scales depend on the parameter choices for the potential and field-space manifold. We provide conclusions and discuss future prospects in Section~\ref{sec:conclusions}.

\section{Models, Parameters, and Initial Conditions}
\label{sec:modeletc}

\subsection{Structure of the Models}

We study inflationary models consisting of multiple real scalar fields $\phi^I$, each nonminimally coupled to the Ricci spacetime curvature scalar. We closely follow the analysis and notation of Refs.~\cite{Kaiser:2012ak,Greenwood:2012aj,Kaiser:2013sna,Schutz:2013fua,DeCross:2015uza,DeCross:2016fdz,DeCross:2016cbs}. We work in $(3+1)$ spacetime dimensions and adopt the ``mostly plus" spacetime metric signature $(-,+,+,+)$. In the Jordan frame, the action takes the form
\beq
S = \int d^4 x \sqrt{-\tilde{g} } \left[ f (\phi^I ) \tilde{R} - \frac{1}{2} \delta_{IJ} \tilde{g}^{\mu\nu} \partial_\mu \phi^I \partial_\nu \phi^J - \tilde{V} (\phi^I ) \right] ,
\label{SJ}
\eeq
where upper-case Latin letters label field-space indices, $I, J = 1, 2, ... , N$, Greek letters label spacetime indices, $\mu, \nu = 0, 1, 2, 3$, and tildes denote Jordan-frame quantities. We will use lower-case Latin letters for spatial indices, $i, j = 1, 2, 3$. 

We bring the gravitational portion of the action into canonical Einstein-Hilbert form by performing a conformal transformation, rescaling $\tilde{g}_{\mu\nu} (x) \rightarrow g_{\mu\nu} (x) = \Omega^2 (x)\> \tilde{g}_{\mu \nu} (x)$ with the function
\beq
\Omega^2 (x) = \frac{2}{M_{\rm pl}^2} f (\phi^I (x)) ,
\label{gtildeg}
\eeq
where $M_{\rm pl} \equiv 1 / \sqrt{8 \pi G} = 2.43 \times 10^{18}$ GeV is the reduced Planck mass. The action in the Einstein frame becomes \cite{Kaiser:2010ps,Abedi:2014mka}
\beq
S = \int d^4 x \sqrt{-g} \left[ \frac{ M_{\rm pl}^2}{2} R - \frac{1}{2} {\cal G}_{IJ} (\phi^K ) g^{\mu\nu} \partial_\mu \phi^I \partial_\nu \phi^J - V (\phi^I ) \right] \, ,
\label{SE}
\eeq 
where the potential is stretched by the conformal factor $f(\phi^I)$,
\beq
V (\phi^I) = \frac{ M_{\rm pl}^4 }{4 f^2 (\phi^I) } \tilde{V} (\phi^I ) .
\label{VE}
\eeq
In addition, the nonminimal couplings induce a curved field-space manifold in the Einstein frame, with associated field-space metric ${\cal G}_{IJ} (\phi^K)$
\beq
{\cal G}_{IJ} (\phi^K ) = \frac{ M_{\rm pl}^2 }{2 f (\phi^K) } \left[ \delta_{IJ} + \frac{3}{ f (\phi^K) } f_{, I} f_{, J} \right] ,
\label{GIJ}
\eeq
where $f_{, I} = \partial f / \partial \phi^I$. For multiple fields ($N \geq 2$), one cannot canonically normalize all of the fields while retaining the Einstein-Hilbert form of the gravitational part of the action \cite{Kaiser:2010ps}.

Varying the action of Eq. (\ref{SE}) with respect to $g_{\mu \nu}$ yields the field equations 
\beq
R_{\mu\nu} - \frac{1}{2} g_{\mu\nu} R = \frac{1}{M_{\rm pl}^2} T_{\mu\nu} ,
\label{EFE}
\eeq
with the energy-momentum tensor given by \cite{Kaiser:2012ak}
\beq
T_{\mu\nu} = {\cal G}_{IJ} \partial_\mu \phi^I \partial_\nu \phi^J - g_{\mu\nu} \left[ \frac{1}{2} {\cal G}_{IJ} g^{\alpha \beta} \partial_\alpha \phi^I \partial_\beta \phi^J + V (\phi^I ) \right] .
\label{Tmn}
\eeq
Varying Eq. (\ref{SE}) with respect to $\phi^I$ yields the equation of motion
\beq
\Box \phi^I + g^{\mu\nu} \Gamma^I_{\> JK} \partial_\mu \phi^J \partial_\nu \phi^K - {\cal G}^{IJ} V_{, J} = 0 ,
\label{eomphi}
\eeq
where $\Box \phi^I \equiv g^{\mu\nu} \phi^I_{\>\> ; \mu  \nu}$ and $\Gamma^I_{\> JK} (\phi^L)$ is the Christoffel symbol constructed from the field-space metric ${\cal G}_{IJ}$.

We numerically simulate Eq.~(\ref{eomphi}) on a lattice for the coupled fields $\phi^I (x^\mu)$. Nonetheless, it is helpful for developing intuition and identifying interesting regions of parameter space to consider a semi-analytical, linearized analysis. For our linearized analysis, we split the fields $\phi^I$ into a background part $\varphi^I$, which satisfies $\nabla \varphi ^I =0$ (where $\nabla$ represents the spatial Laplacian operator) and a fluctuation $\delta \phi^I (x^\mu)$, and work to first order in $\delta \phi^I$. On the lattice such a split is not necessary and the equations are solved for the full fields $\phi^I$. The `background' fields $\varphi^I$ can be obtained from the lattice simulations by averaging the fields over the box size. 

In the linearized analysis, the equation of motion for the fields $\varphi^I$ becomes,
\beq
{\cal D}_t \dot{\varphi}^I + 3 H \dot{\varphi}^I + {\cal G}^{IJ} V_{, J} = 0 ,
\label{eomvarphi}
\eeq
where ${\cal D}_t \equiv \dot{\varphi}^I \, {\cal D}_I$ is the covariant directional derivative with respect to the field-space metric ${\cal G}_{IJ}$ \cite{Gong:2011uw,Kaiser:2012ak,Gong:2016qmq}. Eqs.~(\ref{EFE})-(\ref{Tmn}) yield the usual dynamical equations at background order,
\beq
\begin{split}
H^2 &= \frac{1}{3 M_{\rm pl}^2} \left[ \frac{1}{2} {\cal G}_{IJ} \dot{\varphi}^I \dot{\varphi}^J + V (\varphi^I ) \right] , \\
\dot{H} &= - \frac{1}{2 M_{\rm pl}^2} {\cal G}_{IJ} \dot{\varphi}^I \dot{\varphi}^J .
\end{split}
\label{Friedmann}
\eeq
In Eqs.~(\ref{eomvarphi})-(\ref{Friedmann}), $H \equiv \dot{a} / a$ is the Hubble parameter, and the field-space metric is evaluated at background order, ${\cal G}_{IJ} (\varphi^K)$. In this paper we restrict attention to an unperturbed, spatially flat Friedmann-Lema\^{i}tre-Robertson-Walker (FLRW) spacetime metric.

We consider two-field models with  $\phi^I = \{\phi, \chi\}$, and take $f (\phi^I)$ to be of the form
\beq
f (\phi, \chi ) = \frac{1}{2} \left[ M_{\rm pl}^2 + \xi_\phi \phi^2 + \xi_\chi \chi^2 \right] \, ,
\label{f2field}
\eeq
consistent with renormalization of self-interacting scalar fields in curved spacetime. (Even if the nonminimal couplings $\xi_I$ vanished at background order, terms of the form in Eq.~(\ref{f2field}) would arise from loop corrections \cite{Callan:1970ze,Bunch:1980bs,Bunch:1980br,Birrell:1982ix,Odintsov:1990mt,Buchbinder:1992rb,Parker:2009uva,Markkanen:2013nwa}.) The field-space metric in the Einstein frame, ${\cal G}_{IJ} (\varphi^K)$, in turn, is determined by the form of $f (\phi^I)$ and its derivatives, as in Eq.~\eqref{GIJ}. Explicit expressions for ${\cal G}_{IJ}$ and related quantities for this model may be found in Appendix \ref{sec:AppendixAFieldSpace}.
It was shown in Ref.~\cite{Kaiser:2012ak} that the field-space manifold that arises in the Einstein frame due to the form of  Eq.~\eqref{f2field} is asymptotically flat for large field values, but is strongly curved near the origin $\phi^I=0$; this feature significantly affects preheating for large values of the nonminimal coupling constants \cite{DeCross:2015uza,DeCross:2016fdz,DeCross:2016cbs, Ema:2016dny}.

In this work, we consider a simple, renormalizable form for the potential in the Jordan frame,
\beq
\tilde{V} (\phi, \chi) = \frac{\lambda_\phi}{4} \phi^4 + \frac{g}{2}  \phi^2 \chi^2 + \frac{ \lambda_\chi}{4} \chi^4 .
\label{VJphichi}
\eeq
We take $\lambda_I  > 0$ and neglect bare mass terms ${1\over 2}m_\phi^2\phi^2, {1\over 2}m_\chi^2\chi^2$, in order to focus on effects from the quartic self-couplings and direct interaction terms within a parameter space of manageable size. 
Since  efficient particle production occurs due to the contribution of the field-space curvature near the origin, we do not expect 
effects from nonzero $m_I^2$ to significantly change our results. Effects from nonzero bare masses can be incorporated in our analysis and are left for future work, since they can in principle lead to the production of heavy particles after inflation.
We furthermore focus on positive nonminimal couplings $\xi_I > 0$. (See, e.g., Ref.~\cite{DeCross:2015uza} for a discussion of various constraints on $\xi_I$).

This family of models has been shown to possess strong single-field attractors for large values of $\xi_I$, both during \cite{Kaiser:2013sna} and after inflation \cite{DeCross:2015uza, Nguyen:2019kbm}. Within such an attractor, the fields evolve along a straight trajectory in field-space. 

Inflation begins in a regime in which $\sum_I \xi_I (\phi^I )^2 > M_{\rm pl}^2$. The potential in the Einstein frame becomes asymptotically flat along each direction of field space, as any of the fields $\phi^I$ becomes arbitrarily large. With no loss of generality we can align the field-space coordinate system such that inflation proceeds along the direction $\phi^J$, where the potential reads
\beq
V (\phi^J) \rightarrow \frac{M_{\rm pl}^4}{4} \frac{\lambda_J}{\xi_J^2} \left[ 1 + {\cal O} \left( \frac{ M_{\rm pl}^2}{\xi_J (\phi^J)^2 } \right) \right] ,
\label{Vasympt}
\eeq
(no sum on $J$). Unless some explicit symmetry constrains all coupling constants in the model to be identical ($\lambda_\phi = g = \lambda_\chi$, $\xi_\phi = \xi_\chi$), the potential in the Einstein frame will develop ridges and valleys that satisfy $V > 0$, under the minimal assumption $g>-\sqrt{\lambda_\phi \lambda_\chi}$.

The potential topography was analyzed in Ref.~\cite{Schutz:2013fua} in the limit $\xi_I \gg 1$ and extended to arbitrary $\xi_I>0$ in Ref.~\cite{DeCross:2015uza}. We define the convenient combinations of couplings, which were introduced in Ref.~\cite{Schutz:2013fua},
\beq
\Lambda_\phi \equiv \lambda_\phi \xi_\chi - g \xi_\phi , \>\>\>\> \Lambda_\chi \equiv \lambda_\chi \xi_\phi - g \xi_\chi , \>\>\>\> \varepsilon \equiv \frac{ \xi_\phi - \xi_\chi}{\xi_\phi} ,
\label{Lambdadef}
\eeq
along with the rescaled quantities defined in Ref.~\cite{DeCross:2015uza},
\beq
\tilde{\Lambda}_\phi \equiv \frac{ \Lambda_\phi}{\lambda_\phi \xi_\phi} = \frac{\xi_\chi}{\xi_\phi} - \frac{g}{\lambda_\phi} , \>\>\>\> \tilde{\Lambda}_\chi \equiv \frac{ \Lambda_\chi}{\lambda_\chi \xi_\chi} = \frac{ \xi_\phi}{\xi_\chi} - \frac{g}{\lambda_\chi} .
\label{tildeLambda}
\eeq
(For similar parameterizations, see also Ref.~\cite{Kwapisz:2017vjt}.) For arbitrary $\xi_I \gg 1$ we find ${\cal D}_{ \chi\chi} V \vert_{\chi = 0} \propto - \Lambda_\phi$ and ${\cal D}_{ \phi\phi} V \vert_{\phi = 0} \propto - \Lambda_\chi$ during inflation, elucidating the geometrical interpretation of $\Lambda_{\phi}$ and $\Lambda_{\chi}$ as parameters that characterize the local curvature of the potential perpendicular to the two principal axes.
In the limit $\xi_I\gg 1$, whenever $\Lambda_\phi < 0$ the direction $\chi = 0$ remains a local minimum of the potential and the background dynamics will obey strong attractor behavior along the direction $\chi = 0$. Since the potential of Eq.~\eqref{VJphichi} and the nonminimal coupling function $f$ have two discrete symmetries $\phi \to -\phi$ and $\chi\to -\chi$, we will concentrate only on the positive quadrant $\phi,\chi>0$. The topography of the potential for $\xi_I\gg 1$ is controlled by the parameters $\Lambda_\phi$ and $\Lambda_\chi$, as described in Table~\ref{tab:potentialtopography}. Without loss of generality, we will only consider single-field trajectories that lie along $\chi=0$ for the remainder of this work. The ellipticity parameter $\varepsilon$ describes the morphology of the potential in the case $\Lambda_\phi=\Lambda_\chi=0$, in which case the iso-potential contours are well described by ellipses. More details on the geometrical intuition  regarding the potential parameters of Eq.~\eqref{Lambdadef} and their effects during and after inflation can be found in Refs.~\cite{Schutz:2013fua,DeCross:2015uza}. 

 \begin{table}[h]
\begin{center}
\begin{tabular}{ |c|c|c| } 
 \hline
~ $\Lambda_\phi<0$~ & ~$\Lambda_\chi<0$~ & two single-field attractors along $\phi=0$ and $\chi=0$ \\ \hline
~ $\Lambda_\phi<0$~ & ~$\Lambda_\chi>0$~ & one  single-field  attractor along $\chi=0$ \\ \hline
~ $\Lambda_\phi>0$~ & ~$\Lambda_\chi<0$~ & one  single-field  attractor along $\phi=0$ \\ \hline
~ $\Lambda_\phi>0$~ & ~$\Lambda_\chi>0$~ & one  single-field attractor along $\chi/\phi=\sqrt{\Lambda_\phi / \Lambda_\chi}$ \\ 
 \hline
\end{tabular}
\end{center}
 \caption{\small Potential topography and attractor structure.} 
 \label{tab:potentialtopography}
 \end{table}

Next we must consider observational constraints, such as the present bound on the primordial tensor-to-scalar ratio, $r < 0.1$ \cite{Akrami:2018odb}, which corresponds to the bound $H_* \leq 3.4 \times 10^{-5} \> M_{\rm pl}$. (Asterisks indicate values of quantities at the time during inflation when observationally relevant perturbations first crossed outside the Hubble radius. As described in Ref.~\cite{Akrami:2018odb}, the exact bound on the tensor-to-scalar ratio depends on the combined data sets used, but the resulting uncertainty does not  affect our analysis.) Due to the existence of the strong single field attractor, models in our class predict $r=16\epsilon$ \cite{Kaiser:2012ak,Greenwood:2012aj,Kaiser:2013sna,Schutz:2013fua},
where $\epsilon\equiv -\dot H/H^2$ is the usual slow-roll parameter.

For inflation along $\chi=0$ and $\xi_\phi \gg 1$ we find to good approximation \cite{Kaiser:2013sna}
\beq
H_* \simeq \sqrt{ \frac{\lambda_\phi}{12 \xi_\phi^2 } } \> M_{\rm pl} \, , \quad N_* \simeq \frac{3}{4}\frac{\xi_\phi \phi_*^2}{M_{\rm pl }^2 }\, ,\quad
\epsilon \simeq \frac{3}{4 N_*^2} \, , \quad  \eta \simeq -{1\over N_*} \, ,
\label{Nstarbigxi}
\eeq
where $N_*$ is the number of efolds before the end of inflation when relevant scales crossed outside the Hubble radius. (See also Ref.~\cite{Bezrukov:2013fca}.) Assuming $50 \leq N_* \leq 60$, we find $r \sim {\cal O} (10^{-3})$ in the limit $\xi_\phi \gg 1$, and $H_* \leq 3.4 \times 10^{-5} \> M_{\rm pl}$ for $\lambda_\phi / \xi_\phi^2 \leq 1.4 \times 10^{-8}$. In models like Higgs inflation \cite{Bezrukov:2007ep}, one typically finds $\lambda_\phi \sim {\cal O} ( 10^{-2} - 10^{-4} )$ at the energy scale of inflation (the range in $\lambda_\phi$ stemming from uncertainty in the value of the top-quark mass, which affects the running of $\lambda_\phi$ under renormalization-group flow) \cite{Barvinsky:2009ii,Barvinsky:2009fy,Bezrukov:2012sa,Allison:2013uaa}. That range of $\lambda_\phi$, in turn, requires $\xi_\phi \sim {\cal O} ( 10^2 - 10^3 )$ at high energies  --- a reasonable range, given that $\xi_\phi$ typically rises with energy scale under renormalization-group flow with no UV fixed point \cite{Odintsov:1990mt,Buchbinder:1992rb}. Even for such large values of $\xi_I$, the inflationary dynamics occur at energy scales well below any nontrivial unitarity cut-off scale. (See Ref.~\cite{Schutz:2013fua} and references therein for further discussion.) However, preheating dynamics may involve wavenumbers exceeding the unitarity scale, as was shown to occur in the case of Higgs inflation for $\xi\gtrsim 300$ \cite{Sfakianakis:2018lzf}. In this work we fix $\lambda_\phi / \xi_\phi^2 =  10^{-8}$ and consider nonminimal couplings within the range $1 \leq \xi_I \leq 100$. As discussed in Appendix \ref{sec:AppendixDNumericalConvergence}, accurate simulations of the preheating dynamics in this family of models becomes computationally expensive for $\xi_\phi > 100$.

\subsection{Fluctuations and Parameter Choices}
\label{sec:fluct}

In this subsection we introduce the equations of motion for the linearized perturbations. Although these equations break down in the later stages of reheating due to strong nonlinear effects \cite{Nguyen:2019kbm}, they are useful for understanding the initial parametric resonance as well as for setting initial conditions for our lattice simulations. As noted above, in our lattice simulations we neglect perturbations of the spacetime metric. (The inclusion of metric fluctuations in our lattice simulations remains subject of further study, and could elucidate the limits of linearized gravity during preheating, as has been recently discussed in Ref.~\cite{Giblin:2019nuv}. This would represent a different context of linearized gravity than the one studied recently in Ref.~\cite{Giblin:2018ndw}. Refs.~\cite{Sfakianakis:2018lzf,DeCross:2015uza,DeCross:2016cbs} consider effects of linearized metric perturbations in the early stages of parametric resonance in the types of models we consider here.)

To address the multifield aspects of the models under consideration, we build on the methods reviewed in Refs.~\cite{Wands:2007bd,Gong:2016qmq}
and expand the scalar fields to first order in perturbations, $\phi^I (x^\mu) = \varphi^I (t) + \delta \phi^I (x^\mu)$. For the two-field models considered here, this yields
\beq
\phi (x^\mu) = \varphi (t) + \delta \phi (x^\mu) \, , \qquad \chi (x^\mu) = \delta \chi (x^\mu)\, ,
\label{phivarphi}
\eeq
since the background value of the $\chi$ field at lowest order is exponentially close to zero, due to the single-field attractor during inflation. To first order in $\delta\phi^I$, Eqs.~(\ref{EFE})-(\ref{eomphi}) may be combined to yield the equation of motion for the perturbations \cite{Langlois:2008mn,Kaiser:2012ak,Renaux-Petel:2015mga} 
\beq
{\cal D}_t^2 {\delta\phi}^I + 3 H {\cal D}_t {\delta\phi}^I + \left[ \frac{k^2}{a^2} \delta^I_{\> J} + {\cal M}^I_{\> J}  \right] {\delta\phi}^J = 0 ,
\label{eomQ}
\eeq
where the mass-squared tensor takes the form
\beq
{\cal M}^I_{\> J} \equiv {\cal G}^{IK} \left( {\cal D}_J {\cal D}_K V \right) - {\cal R}^I_{\> LMJ} \dot{\varphi}^L \dot{\varphi}^M\label{MIJ}\, ,
\eeq
and ${\cal R}^I_{\> LMJ}$ is the Riemann tensor for the field-space manifold. All expressions in Eqs.~(\ref{eomQ}) and (\ref{MIJ}) involving ${\cal G}_{IJ}$, $V$, and their derivatives are evaluated at background order in the fields, $\varphi^I$. 
Compared with the corresponding expression for ${\cal M}^I_{\> J}$ in Ref.~\cite{DeCross:2015uza}, the expression in Eq.~(\ref{MIJ}) is missing a term proportional to $1/M_{\rm pl}^2$. That term arises from the coupled metric perturbations, which we neglect in the present analysis.

Due to the existence of a single-field attractor for the motion of the background fields, at least in the linearized regime in which one neglects backreaction effects, the mass matrix ${\cal M}^I_{\> J}$ becomes diagonal and the equations of motion for $\delta\phi$ and $\delta\chi$ decouple \cite{DeCross:2015uza}. We rescale the fluctuations $\delta\phi^I (x^\mu) \to X^I (x^\mu) / a(t)$, promote the  $X^I$  to operators $\hat X^I$, and quantize $\hat X^\phi$ and $\hat X^\chi$ by expanding each in sets of creation and annihilation operators and associated mode functions. Within a single-field attractor, the resulting expansion simplifies to \cite{DeCross:2015uza}
\begin{eqnarray}
\hat X^\phi (x^\mu) &=& \int {d^3k \over (2\pi)^{3/2}} \sqrt{{\cal G}^{\phi\phi} (t) } \left [
v_k (t) \, \hat b_{\bf k} e^{i {\mathbf k}\cdot {\mathbf x}} +v_k^* (t) \, \hat b_{\bf k}^\dagger e^{-i {\mathbf k}\cdot {\mathbf x}}  
\right ], \label{Xphi}
\\
\hat X^\chi (x^\mu) &=& \int {d^3k \over (2\pi)^{3/2}} \sqrt{{\cal G}^{\chi\chi} (t)} \left [
z_k (t) \, \hat c_{\bf k} e^{i {\mathbf k}\cdot {\mathbf x}} +z_k^* (t) \, \hat c_{\bf k} ^\dagger e^{-i {\mathbf k}\cdot {\mathbf x}}  
\right ] \, , \label{Xchi}
\end{eqnarray}
where $[ \hat{b}_{\bf k} , \hat{b}^\dagger_{\bf q} ] = [ \hat{c}_{\bf k} , \hat{c}^\dagger_{\bf q} ] = \delta^{(3)} ({\bf k} - {\bf q} )$, and all other commutators among $\{ \hat{b}_{\bf k} , \hat{b}_{\bf k}^\dagger , \hat{c}_{\bf k} , \hat{c}^\dagger_{\bf k} \}$ vanish.
The linearized equations of motion for the mode functions $v_k(t)$ and $z_k(t)$ for the $\phi$ and $\chi$ fluctuations, respectively, become
\begin{subequations}
\label{eq:flucteom}
\begin{eqnarray}
\ddot v_k + H\dot v_k + \Omega^2_\phi v_k =0,
\\
\ddot z_k + H\dot z_k + \Omega^2_\chi z_k=0,
\end{eqnarray}
\end{subequations}
where
\beq
\Omega_I^2 = {k^2\over a^2} + m_{{\rm eff},I}^2,
\label{OmegaIdef}
\eeq
and the effective mass-squared for the fluctuations (within the single-field attractor) is given by
\begin{eqnarray}
m_{{\rm eff},\phi}^2 &=& {\cal G}^{\phi\phi}{\cal D}_\phi \partial_\phi V -{1\over 6}R \, ,
\\
m_{{\rm eff},\chi}^2 &=& {\cal G}^{\chi\chi}{\cal D}_\chi \partial_\chi V  -{\cal R}^\chi_{~\phi\phi\chi}  \dot\varphi^2 -{1\over 6}R \, .
\end{eqnarray}
The quantity $R=6(2-\epsilon)H^2$ is the space-time Ricci scalar. The term proportional to $R$ in both $m_{{\rm eff},\phi}^2$ and $m_{{\rm eff},\chi}^2$ remains subdominant during preheating, as shown in Ref.~\cite{DeCross:2015uza}. We can characterize the dominant contributions to the effective mass of the fluctuations in terms of $\Lambda_\phi$ and $\varepsilon$, defined in Eq.~\eqref{Lambdadef}. For trajectories that proceed along $\chi=0$ the quantity $\Lambda_\chi$ does not enter the fluctuation analysis at the linear level.

Inflation ends at $\xi_\phi \varphi^2 ={\cal O}(1) M_{\rm pl}^2$, so we define the rescaled field amplitude $\delta(t) \equiv \sqrt{\xi_\phi} \varphi(t)/M_{\rm pl}$ to study the preheating dynamics. In the regime $\xi_\phi\gg1$ the effective mass of the $\phi$ fluctuations is dominated by the second derivative of the potential,
\begin{equation}
{\cal G}^{\phi\phi}{\cal D}_\phi {\cal D}_\phi V \simeq
{
\lambda_\phi  M_{\rm pl}^2 \over \xi_\phi
} 
\frac{\delta ^2  \left(3 -2\xi_\phi \delta ^4+2\xi_\phi \delta ^2 \right)}{\left(\delta ^2+1\right)^2 \left(6 \delta ^2+1\right) \left(1+ \xi_\phi  \delta^2\right)} \, .\label{eq:masschi}
\end{equation}
The corresponding potential contribution for the $\chi$ fluctuations in the regime $\xi_I\gg 1$ is
\beq
{\cal G}^{\chi\chi}{\cal D}_\chi {\cal D}_\chi V \simeq 
  \begin{cases} 
- 6 \tilde \Lambda_\phi \,\lambda_\phi \, M_{\rm pl}^2 \, { \delta^4  \over
(1+\delta^2) (1+6\xi_\phi\delta^2)} \, ,
  & \quad \tilde \Lambda_\phi \ne 0 \, , \\
{g\over \lambda_\phi \xi_\phi} \, \lambda_\phi \, M_{\rm pl}^2 \, { \delta^2 \over
(1+\delta^2)^2 (1+6\xi_\phi\delta^2)}\, ,  & \quad  \tilde \Lambda_\phi= 0\, .\
\end{cases}
\label{eq:DchichiV}
\eeq
We can distinguish between the two cases in Eq.~\eqref{eq:DchichiV} by considering the pre-factors, $6\tilde \Lambda_\phi$ and $g/(\lambda_\phi \xi_\phi)$, the latter valid for $\tilde \Lambda_\phi=0$. Since generically $\tilde \Lambda_\phi ={\cal O}(1)$, $g/\lambda_\phi={\cal O}(1)$ and $\xi_\phi \gg 1$, the potential contribution to the effective mass is proportional to $\tilde \Lambda_\phi = {\cal O}(1)$ for a generic choice of parameters. In the symmetric case ($ \Lambda_\phi= 0)$ the potential contribution to the $\chi$ mass is significantly reduced, since it is proportional to $1/\xi_\phi \ll 1$. 

The contribution of the field-space structure enters through the Riemann term ${\cal R}^\chi_{~\phi\phi\chi}$ and the background field velocity $\dot \varphi(t)$. Since $\dot\varphi(t)$ only depends on $\xi_\phi$ and not on $\tilde\Lambda_\phi$ and $\varepsilon$ within the single-field attractor, we will analyze the Riemann term, which can be approximated by 
\beq
{\cal R}^\chi_{~\phi\phi\chi} \simeq -\frac{ 6 \xi_\phi^2 }{M_{\rm pl}^2} (1-\varepsilon) \, ,
\eeq
when the $\varphi$-field crosses zero. When the inflaton field crosses the origin at $\delta(t) =0$, both the field-space curvature and the field velocity are maximized. Hence the height of the ``Riemann spike'' \cite{DeCross:2015uza} depends crucially on the value of the ellipticity. For large values of the nonminimal couplings $\xi_I\gg1$, the Riemann spike is well described (around its maximum) by a Lorentzian function
\beq
{\cal R}^\chi_{~\phi\phi\chi} \simeq -\frac{ 6 \xi_\phi^2 }{M_{\rm pl}^2 } \frac{ (1-\varepsilon)}{ ( 1+6\xi_\phi \delta^2 )}.
\label{eq:lorentzian}
\eeq
Interestingly, while the magnitude of the Riemann spike grows for large negative values of the ellipticity $\varepsilon<0$, the Riemann term appears both in the numerator and denominator of the adiabaticity parameter $\cal A_\chi$ (see Ref.~\cite{DeCross:2015uza}),
\beq
{\cal A}_\chi = {\partial_t m_{{\rm eff},\chi}^2\over 2m_{{\rm eff},\chi}^3} + {H\over m_{{\rm eff},\chi}} +{\cal O}\left(
{k^2\over (aH)^2}
\right )  \, .
\eeq
It can be shown (see Appendix \ref{sec:AppendixBAdiabaticity}) that, at least for the first few inflaton zero-crossings, 
\beq
{\cal A}_\chi \propto  {1\over \sqrt{1-\varepsilon} } \, ,
\eeq
where the proportionality factor is about $1.8$. Thus, in the limit of large nonminimal couplings, for a given value of $\tilde \Lambda_\phi$, a larger Riemann spike (due to a negative ellipticity) can actually lead to a suppression of parametric resonance, by lowering the non-adiabaticity parameter. This agrees with the Floquet analysis of Ref.~\cite{DeCross:2016fdz}, in which altering the ellipticity was shown to significantly affect the instability chart for $\chi$ fluctuations. 

Based on these intuitive relations between the effective mass of fluctuations and the potential topography parameters, we choose a specific set of couplings that capture all characteristic cases, summarized in Table \ref{tab:benchmarks}. These five cases extend the analysis in Ref.~\cite{Nguyen:2019kbm}. For easy comparison, all cases except for the symmetric one (B) have $\tilde \Lambda_\phi=-0.2$, only one minimum per quadrant ($\tilde \Lambda_\chi>0$), and a fixed ratio of the potential height at the two extrema: 
\beq
{V(\varphi=0, \chi\to \infty)\over V(\varphi\to \infty,\chi=0)} = 0.512 \, .
\eeq
We consider Case A (the ``benchmark") to be fairly generic for this family of models, and study the preheating dynamics of that case in detail in Section \ref{sec:benchmark}. In Section \ref{sec:parameters} we highlight how the characteristic time-scales for various nonlinear processes shift as we change the parameters for Cases B-E.

 \begin{table}[h]
\begin{center}
\begin{tabular}{ |c|c|c|c|c|c|} 
 \hline
A& $\xi_\chi=0.8\xi_\phi$ & $g=\lambda_\phi $ &$\lambda_\chi=1.25 \lambda_\phi$  & benchmark case\\ \hline
B& $\xi_\chi=\xi_\phi$ & $g= \lambda_\phi $ &$\lambda_\chi = \lambda_\phi$ & symmetric \\ \hline
C& $\xi_\chi=2\xi_\phi$ & $g=2.2 \lambda_\phi $ &$\lambda_\chi\simeq 7.8 \lambda_\phi$ & negative ellipticity \\ \hline
D& $\xi_\chi=0.5\xi_\phi$ & $g=0.7 \lambda_\phi $ &$\lambda_\chi\simeq 0.49 \lambda_\phi$  & positive ellipticity\\ \hline
E& $\xi_\chi=\xi_\phi$ & $g=1.2 \lambda_\phi $ &$\lambda_\chi\simeq 1.95 \lambda_\phi$  & zero ellipticity \\ \hline
\end{tabular}
\end{center}
 \caption{\small Parameter choices for the five characteristic cases.} 
 \label{tab:benchmarks}
 \end{table}

\subsection{Initial Conditions for Lattice Simulations}

Initial conditions for the preheating simulations can be set by applying the Wentzel-Kramers-Brillouin (WKB) approximation to the equations of motion for the fluctuations $v_k$ and $z_k$, written in terms of conformal time $d\eta = dt/a$ as
\beq
\partial_\eta^2 v_k + { \Omega^2_{\phi } (k, \eta)} \, v_k =0 ,
\eeq
and similarly for $z_k$ with frequency $\Omega_\chi (k, \eta)$. The WKB analysis is valid as long as the adiabaticity condition is satisfied,
$\Omega'_{I}/\Omega^2_{I} \ll1 $, where a prime denotes $d / d\eta$. The mode functions for the $\phi$ fluctuations thus become
\beq
v_k(\eta) = {1\over \sqrt{2\Omega_{\phi} (k, \eta) }} e^{-i \int d\eta' \, \Omega_{\phi} (k,\eta')} .
\eeq
Since the lattice code computes field values $\phi(t, {\bf x})$ and $\chi(t, {\bf x})$, we can easily use Eq.~(\ref{Xphi}) to relate $\delta\phi_k$ to $v_k$ as
\beq
\delta\phi_k = {1\over a} \sqrt{{\cal G}^{\phi\phi}} \,v_k
 ={ \sqrt{{\cal G}^{\phi\phi}}\over \sqrt{2\Omega_{\phi} (k,\eta) }} e^{-i \int d\eta' \, \Omega_{\phi} (k,\eta')} ,
 \label{deltaphiinit}
\eeq
where we have normalized the scale-factor to unity at the end of inflation. The time derivative is then given by 
\beq
\partial_t  {\delta\phi}_k =\left (-H + {1\over 2}  {\partial_t {\cal G}^{\phi\phi}\over {{\cal G}^{\phi\phi}}} +
{\partial_t \Omega_{\phi} \over \Omega_{\phi} }-i\Omega_{\phi} 
 \right )\delta\phi_k \, ,
 \label{deltaphidotinit}
\eeq
which can be evaluated at the initial time to fix the initial conditions for the $\phi$ fluctuations. Following similar steps, the initial conditions for the $\chi$ fluctuations are given by
\begin{eqnarray}
\delta\chi_k  &=& \sqrt{{\cal G}^{\chi\chi}}\over \sqrt{2\Omega_{\chi} (k,\eta) }} e^{-i \int d\eta' \, \Omega_{\chi} (k,\eta') \, , \label{deltachiinit}
\\
\partial_t  {\delta\chi}_k &=&\left (-H + {1\over 2}  {\partial_t {\cal G}^{\chi\chi}\over {{\cal G}^{\chi\chi}}} +
{\partial_t \Omega_{\chi} \over \Omega_{\chi} }-i\Omega_{\chi}
 \right )\delta\chi_k \, . \label{deltachidotinit}
\end{eqnarray}
We see that the contribution of the non-trivial field-space metric is especially important for the initial magnitude of the fluctuations and their relative size, since at the end of inflation 
\beq
{\cal G}^{\phi\phi}  \sim \frac{1}{\xi_\phi} \, ,\quad {\cal G}^{\chi\chi} \sim {\cal O} (1) .
\eeq
The relative magnitude of the two fluctuations at the start of preheating is vastly different for large values of $\xi_\phi$.

We employ a modified version of GABE (Grid and Bubble Evolver) \cite{GABE,Child:2013ria} to evolve the fields and the background, according to Eqs.~(\ref{eomphi}) and (\ref{Friedmann}). We start the simulations when inflation ends, defined by $\epsilon (t_{\rm init}) = 1$; the Hubble scale at this time is $H_{\rm end}$ (referring to the end of inflation). We use a grid with ${\cal N} = 256^3$ points and a comoving box size $L = \pi / H_{\rm end}$, so that the longest wavelength in our spectra corresponds to $k = H_{\rm end} / 2$. We match the two-point correlation functions of $\phi (t_{\rm init}, {\bf x})$ and $\chi (t_{\rm init}, {\bf x})$ to corresponding distributions for quantized field fluctuations, based on Eqs.~(\ref{deltaphiinit}) - (\ref{deltachidotinit}). The initial spectra of the fields are subject to a window function that suppresses high-momentum modes above some UV suppression scale, $k_{\rm UV} = 50 \, H_{\rm end}$. As we discuss in Appendix \ref{sec:AppendixDNumericalConvergence}, the late-time numerical results are largely insensitive to varying $k_{\rm UV}$ between $25 \, H_{\rm end}$ and $100 \, H_{\rm end}$, or changing our grid size from ${\cal N} = 256^3$ points to ${\cal N} = 512^3$ points.

\section{Distinct Processes and Time-Scales during Preheating}
\label{sec:benchmark}

As noted in Section~\ref{sec:fluct}, parametric resonance in this family of models is governed by the effective masses, $m_{{ \rm eff}, I} (t)$, the dominant contributions to which are qualitatively different for the two fields. The self-resonance of the inflaton field $\phi$ is governed by the (covariant) curvature of the potential, while the effective mass of the $\chi$ modes receives an extra contribution from the curved field-space manifold. This can lead to very efficient preheating \cite{Nguyen:2019kbm}, while also exhibiting an interesting interplay between the two dominant contributions to $m_{\rm eff, \chi}$ \cite{DeCross:2015uza,DeCross:2016fdz}. 

\subsection{Time-scales}

\begin{figure}
\centering
\includegraphics[width=0.65\textwidth]{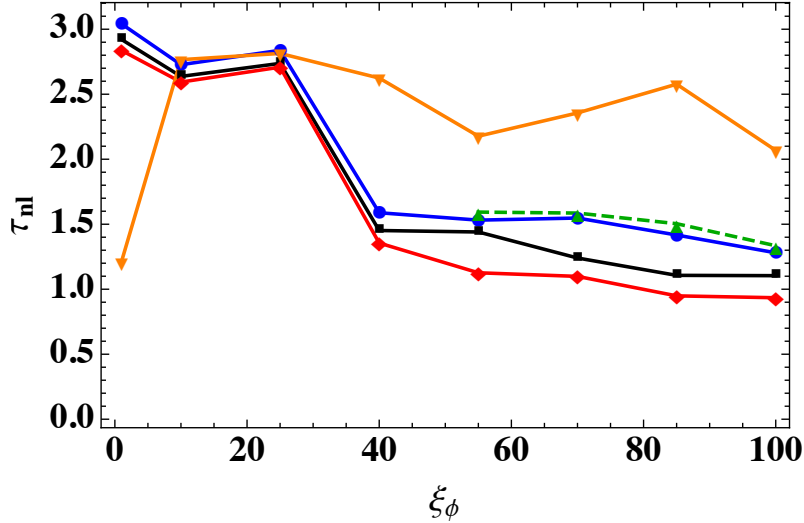}
\caption{ \small \baselineskip 11pt
Nonlinearity time scales $\tau_{\rm nl}$ in $e$-folds after the end of inflation as a function of the nonminimal coupling $\xi_\phi$ for the benchmark Case A: $\tau_{{\rm bg},0.9}$ (blue), $\tau_{\chi_{\rm rms},0.5}$ (black), $\tau_{\phi_{\rm rms},5}$ (red), $\tau_{\rho,0.95}$ (dashed green), and $\tau_{{\rm eos},1/30}$ (orange).  }
 \label{fig:timescalea}
\end{figure}

To quantify the efficiency of preheating, we introduce several time-scales that characterize distinct nonlinear processes after the end of inflation:
\begin{enumerate}
    \item $\tau_{\rm bg}$: The time at which the backreaction of the produced fluctuations starts to affect the evolution of the background field.
    \item $\tau_{\chi_{\rm rms}}$: The time at which the root-mean-square (RMS) value of the $\chi$ fluctuations becomes comparable to the background inflaton value.
    \item $\tau_{\phi_{\rm rms}}$: The time at which fluctuations in the $\phi$ field start growing as a result of nonlinear interactions with the produced $\chi$ fluctuations. 
    \item $\tau_\rho$: The time at which the fluctuations account for  a significant fraction of the total energy density. 
    \item $\tau_{\rm eos}$: The time at which the universe starts expanding in an (approximately) radiation-dominated manner.
\end{enumerate}
In this section we examine each of these time-scales for the ``benchmark" case (Case A), with ratios of parameters given in Table \ref{tab:potentialtopography}. The results are summarized in Fig.~\ref{fig:timescalea}, in which we can distinguish three relevant parameter regimes: $\xi_\phi \sim 1, 10, 100$. These regimes display different physical behavior because of different trade-offs between dominant contributions to $m_{{ \rm eff}, I}$, consistent with the analysis of the linearized perturbations in  Ref.~\cite{DeCross:2016fdz}. We discuss the full nonlinear dynamics of the different regimes below. In all plots the field values are shown in units of the reduced Planck mass, $M_\text{pl}$.

\subsubsection{{$\tau_{\rm bg}$: backreaction on the background field}}

Several recent studies of preheating in nonminimally coupled models \cite{DeCross:2015uza,DeCross:2016fdz,DeCross:2016cbs, Ema:2016dny} and its applications to Higgs preheating \cite{Sfakianakis:2018lzf, Ema:2016dny} were based on linearized analyses, working to first order in the field fluctuations $\delta \phi^I$. In those studies, complete preheating was taken to be the time when the energy density of the (linearized) fluctuations became equal to the energy density of the background field. However, once a significant number of particles has been produced, the backreaction on the background field can no longer be neglected; that is, the decay of the lattice-averaged field $\langle \phi \rangle$ cannot be described by gravitational redshift alone. We define the timescale $\tau_{{\rm bg},x}$ as the moment when $\langle \phi \rangle$ (as computed on the lattice) becomes smaller than $ x\cdot \varphi(t)$, where $\varphi$ is computed in the linearized analysis (which neglects backreaction, so that $\varphi (t)$ only decreases due to redshift). Here $x$ is a dimensionless parameter that encodes the amount of divergence between the (nonlinear) lattice and the (linearized) background evolution. This time-scale is indicated by the vertical blue line in Fig.~\ref{fig:illustratetimescalesA} for $\xi_\phi = 1$ and $\xi_\phi= 100$, for $x=0.9$.

\subsubsection{$\tau_{\chi_{\rm rms}}$: $\chi_{\rm rms}$ becomes comparable to $\langle \phi \rangle$}

 During preheating, parametric resonance drives the production of quanta with specific wavenumbers. Despite the inherent complexity of the underlying resonance structure,
 the number density of produced $\phi^I$-quanta can be quantified by the root-mean-square (RMS) value of the $\phi^I$-field on the lattice: $\phi^I_{\rm rms} \equiv \sqrt{\langle (\phi^{I})^2\rangle -\langle\phi^I\rangle^2}$. In the models considered here, the production of $\chi$-quanta is typically much more efficient than the production of $\phi$-quanta, because of the Riemann spike. We define the time-scale $\tau_{\chi_{\rm rms},x}$ as the moment when the RMS value of the $\chi$-field becomes larger than $x\cdot \langle \phi \rangle $. In Fig.~\ref{fig:illustratetimescalesA}, the vertical black line indicates $\tau_{\chi_{\rm rms},0.5}$ for $\xi_\phi = 1$ and $\xi_\phi= 100$.

 \subsubsection{$\tau_{\phi_{\rm rms}}$: $\phi_{\rm rms}$ grows as a result of rescattering}
 
\begin{figure}
\centering
\includegraphics[width=0.95\textwidth]{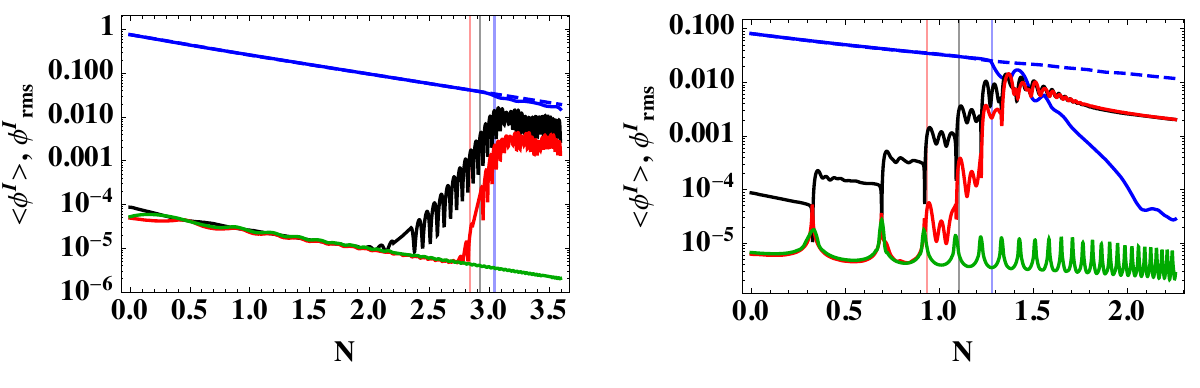}
\caption{\small \baselineskip 11pt
The behavior of various quantities in $e$-folds $N$ after the end of inflation, for Case A: the lattice-averaged background field $\langle \phi \rangle$ (blue); the background field in the linearized analysis $\varphi$ (dashed blue); $\chi_{\rm rms}$ from the lattice (black); $\phi_{\rm rms}$ from the lattice (red); and $\phi_{\rm rms}$ from the linearized analysis (green). All fields are plotted in units of the reduced Planck mass. The values of the nonminimal couplings are $\xi_\phi=1$ (left) and $\xi_\phi = 100$ (right). The vertical blue, black, and red lines indicate $\tau_{{\rm bg},0.9}$, $\tau_{\chi_{\rm rms},0.5}$ and $\tau_{\phi_{\rm rms},5}$, respectively.
 }
\label{fig:illustratetimescalesA}
\end{figure}

 As the self-resonance in the $\phi$-field is typically much weaker than the resonance in the $\chi$ field (which is driven by the field-space curvature), rescattering of $\chi$-modes is the dominant mechanism for amplifying $\phi$-modes. This mechanism is not present in the linearized analysis, so we can quantify the strength of this effect by comparing $\phi_{\rm rms}$ as computed using the lattice results and using the linearized analysis (which neglects backreaction). The quantity $\tau_{\phi_{\rm rms},x}$ is defined as the moment when $\phi_{\rm rms}$ as computed on the lattice exceeds $x\cdot \phi_{\rm rms}$ in the linearized analysis. This time-scale does not depend solely on the efficiency of production, but also on the coupling between $\phi$- and $\chi$-modes. The time $\tau_{\phi_{\rm rms},5}$ is indicated by the vertical red line in Fig.~\ref{fig:illustratetimescalesA}.

\subsubsection{$\tau_\rho$: significant energy transfer}

\begin{figure}
\centering
\includegraphics[width=0.95\textwidth]{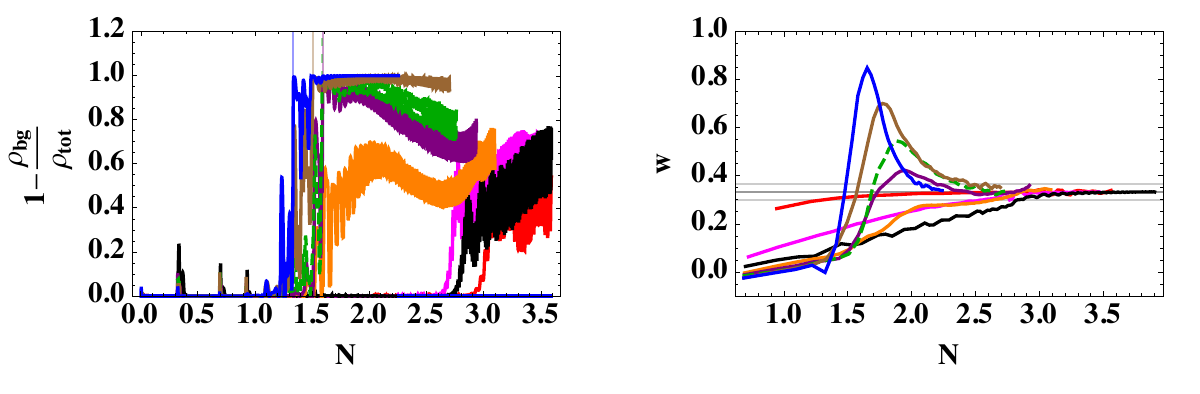}
\caption{\small \baselineskip 11pt
{\it Left:} The fraction of the energy density in produced particles $[1- (\rho_{\rm bg} / \rho_{\rm tot})]$ for Case A versus $e$-folds $N$ after the end of inflation, for $\xi_\phi = 1, 10, 25, 40, 55, 70, 85, 100$ (red, magenta, black, orange, purple, dashed green, brown, blue, respectively). The  vertical lines indicate $\tau_{\rho,0.95}$ for $\xi_\phi = 55,70,85, 100$, in the color corresponding to the appropriate curve.  
{\it Right:} Equation of state for Case A, averaged over one inflaton period, for $\xi_\phi = 1, 10, 25, 40, 55, 70, 85, 100$ (same color coding as in the left plot).
}
\label{fig:Arho1100}
\end{figure}

The fraction of energy density in the produced particles is an important measure of the efficiency of preheating. The time-scale $\tau_{\rho,x}$ is defined as the moment when $[1 - (\rho_{\rm bg}/ \rho_{\rm tot}) ] > x$. The plot of $\tau_{\rho,0.95}$ in Fig.~\ref{fig:timescalea} demonstrates that only for $\xi_\phi \geq 55$ are the preheating processes efficient enough to transfer as much as $95\%$ of the energy density from the background into a bath of produced particles. Fig.~\ref{fig:Arho1100} shows $[1 - (\rho_{\rm bg} / \rho_{\rm tot})]$ as a function of time for $\xi_\phi =1, 10, 25, 40, 55, 70, 85, 100$ (red, magenta, black, orange, purple, dashed green, brown, blue, respectively). The time $\tau_{\rho,0.95}$ is indicated for $\xi_\phi = 55,70, 85, 100$. 

In the case of preheating with a simple massive inflaton field, if some energy density remains in the inflaton condensate after the phase of parametric resonance, the inflaton will eventually become the dominant constituent again, because of the faster rate at which energy density redshifts in fluctuations ($\sim a^{-4} (t)$) compared to that of the inflaton condensate ($\sim a^{-3} (t)$). We see a similar result for $\xi_\phi=55, 70$. There the  fluctuations grow by draining over $90\%$ of the energy density of the inflaton, without completely preheating the universe. After $N\simeq 1.5$ $e$-folds we see that the fractional energy in fluctuations falls. Whereas the quartic potential of the inflaton field close to the origin in the family of models we consider here might suggest that the condensate and fluctuation would redshift at the same rate ($\sim a^{-4} (t)$), two effects lead to a growth of $\rho_{\rm bg}/\rho_{\rm tot}$ and a departure from complete preheating. For nonminimal couplings $\xi>{\cal O}(10)$, the background evolves with an equation of state close to $w=0$ \cite{DeCross:2015uza}. Furthermore, the right panel of Fig.~\ref{fig:Arho1100} shows that during the initial growth of fluctuations, they acquire a stiff equation of state $w>1/3$ for $1.5\lesssim N\lesssim 2.5 $ for $\xi_\phi=55,70$. Hence, even if the energy density of the condensate redshifts as $a^{-4}(t)$, the fluctuations temporarily redshift even faster.

\subsubsection{$\tau_{\rm eos}$: emergence of radiation-dominated equation of state}
 
One of the most significant potential impacts of the reheating era on observables is the evolution of the scale factor as the universe transitions from inflation (with an equation of state $w\simeq -1$) to radiation-dominated expansion (with $w=1/3$). Variations in the expansion history after inflation, encoded through time-varying $w$, affect the matching between CMB-relevant modes and the time when they first exited the Hubble radius during inflation \cite{Martin:2010kz,Mielczarek:2010ag,Martin:2014nya,Munoz:2014eqa,Creminelli:2014oaa,Creminelli:2014fca,Dai:2014jja,Cai:2015soa,Cook:2015vqa,Eshaghi:2016kne,Ueno:2016dim,Figueroa:2018twl}. In the case of a minimally coupled, massless inflaton  field, like one governed by a quartic potential $V(\phi)\propto \phi^4$, the oscillations of the inflaton condensate correspond to $w = 1/3$ immediately after the end of inflation, even before the production of significant numbers of higher-momentum particles. Whereas for small values of the nonminimal coupling $\xi_\phi$, the background evolution after inflation is similar to the minimally coupled case \cite{DeCross:2016fdz}, the behavior is significantly different for $\xi_\phi \gg 1$, for which the usual virial-theorem calculation for a scalar field must be modified to include a contribution from the field-space metric \cite{Kaiser:2013sna}. As $\xi_\phi$ is increased, the oscillating inflaton condensate spends an increasingly long time exhibiting a matter-like equation of state, as described in detail in Ref.~\cite{DeCross:2015uza}. In that same regime of parameter space ($\xi_\phi \gg 1$), however, higher-momentum modes of the $\chi$-field can be efficiently produced, which will transfer energy from the inflaton condensate into a bath of radiative degrees of freedom, eventually yielding a radiation-dominated equation of state. We define the time-scale $\tau_{\rm eos}$ as the time after which the equation of state of the $\{\phi,\chi\}$ system remains close to $w=1/3$. More precisely, we define $\tau_{{\rm eos},x}$ as the time after which $|w-1/3|\le x$, until the end of our simulations, regardless of whether $w$ is mostly controlled by the background field $\varphi$ or by the produced $\phi$ and $\chi$ particles.


 \subsection{Dynamics and Competition of Mass Scales}

We found in the previous subsection that the universe transitions to a radiation-dominated expansion ($w\simeq 1/3$) within fewer than 3 $e$-folds, for $\xi_\phi$ in the range $[1,100]$. However, complete preheating --- as indicated by the transfer of energy from the inflaton condensate into radiative degrees of freedom --- only occurs for large nonminimal couplings, close to $\xi_\phi \sim 100$. Based on this behavior, combined with the linearized analysis of Ref.~\cite{DeCross:2016fdz}, we can distinguish three regimes, based on the value of the nonminimal coupling $\xi_\phi$.

\subsubsection{Small-coupling regime}

We start with a value of the nonminimal coupling $\xi_\phi \sim 1$, in the small-coupling regime \cite{DeCross:2015uza}. 
Fig.~\ref{fig:illustratetimescalesA} shows the evolution of the spatially averaged  value of the inflaton $\langle \phi \rangle$, as well as the RMS values of both fields. Initially only $\chi_{\rm rms}$ grows; $\phi_\text{rms}$ starts to grow in tandem with $\chi_\text{rms}$ as soon as the latter grows beyond $\chi_\text{rms} \approx 10^{-3} M_\text{pl}$. We see that $\langle \phi \rangle$ remains larger than the RMS values of each field, and that the expression for $\varphi$ from the linearized treatment remains an excellent approximation until 3 $e$-folds after the end of inflation. Furthermore, the averaged value of the $\chi$-field, $\langle \chi \rangle$, remains much smaller than $\langle \phi \rangle$ for the entire duration of the oscillation, confirming that the single-field attractor persists during reheating. (We consider the evolution of the covariant turn-rate below.)

The left panel of Fig.~\ref{fig:Arho1100} shows the fraction of the total energy density in radiation modes for a range of values of $\xi_\phi$. Until $N = 2.7$ $e$-folds after the end of inflation, the energy density is almost entirely in the background field for $\xi_\phi = 1$. Towards the end of the simulation, at $N = 3.5$ $e$-folds after the end of inflation, significant energy density has been transferred to the fluctuations, but the energy density in the inflaton condensate remains sizeable, indicating that preheating does not complete for $\xi = 1$. Since the energy density in both the fluctuations and in the background redshifts approximately as $a^{-4} (t)$ for $\xi_\phi \sim 1$, additional couplings are needed to completely reheat the universe (perturbatively) in this case, for example a mass term for the inflaton field $\phi$ and a cubic coupling $\phi\chi^2$, to allow for pair-production processes $\phi \to \chi+\chi$.

\begin{figure}
\centering
\includegraphics[width=0.95\textwidth]{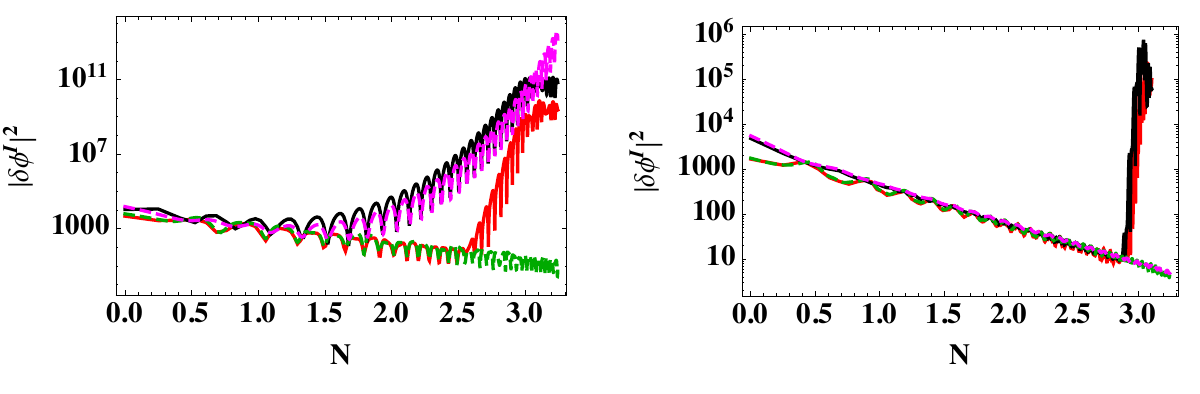}
\caption{\small \baselineskip 11pt
Growth of $|\delta \phi_k|^2$ and $|\delta \chi_k|^2$ for Case A in units of $M_{\rm pl}$ for $k \approx 3 H_{\rm end}$ (left) and $k \approx 11 H_{\rm end}$ (right), with $\xi_\phi = 1$. The red (black) line shows the evolution of $|\delta\phi_k|^2$ ($|\delta\chi_k|^2$) on the lattice, and the green (magenta) dashed lines show the corresponding results in the linearized analysis.
}
\label{fig:ModebyModexi1}
\end{figure}

We can account for the behavior shown in Fig.~\ref{fig:Arho1100} by investigating the growth of $\phi_\text{rms}$ and $\chi_\text{rms}$ in more detail. Fig.~\ref{fig:ModebyModexi1} shows the evolution of the modes $k \approx 3 \, H_{\rm end}, 11 \, H_{\rm end}$  as computed on the lattice and in a linearized treatment. (We discuss how to make such wavenumber-specific comparisons of lattice and linearized treatments in Appendix \ref{sec:AppendixCLatticeCompare}.) In the left panel, we see a weak parametric resonance of the $\chi$ fluctuations for small wavenumbers around $k=3H_{\rm end}$. This can be explained from the behavior found in Ref.~\cite{DeCross:2016fdz}, in which the Fourier decomposition of the background field $\varphi(t)$ and the Floquet structure of the resonances were discussed (to linear order in fluctuations). It was found that the Fourier modes of $\varphi(t)$ collapse to a pure sinusoidal oscillation for $\xi_\phi \simeq 1$. For that same regime of $\xi_I$, the instability bands of the $\chi$ fluctuations are significantly reduced in width (``pinched''), thereby leading  to a set of very narrow instability bands. This is exactly mirrored by the behavior in the left panel of Fig.~\ref{fig:ModebyModexi1}, in which the growth in $|\delta \chi_k|^2$ is caused by a narrow resonance at $k \approx 3 H_\text{end}$. At $2.5$ $e$-folds, $|\delta \phi_k|^2$ starts to grow as a result of rescattering of $\delta \chi_k$-modes into $\delta \phi_k$-modes, an effect that is not captured by the linearized analysis. The right panel of Fig.~\ref{fig:ModebyModexi1} shows that modes with $k \approx 11 H_\text{end}$ start to grow somewhat later, at 2.8 $e$-folds, solely as a result of rescattering. From Fig.~\ref{fig:illustratetimescalesA} we see that backreaction of the produced particles on $\langle\phi\rangle$ causes a deviation from the linearized treatment by $\tau_{\rm bg} \simeq 3$ $e$-folds after the end of inflation for $\xi_\phi \simeq 1$, which is consistent with the moment at which the resonance in $\delta \chi_k$ shuts off in the left panel of Fig.~\ref{fig:ModebyModexi1}.

\subsubsection{Intermediate-coupling regime}
\label{sec:intermediate}

\begin{figure}
\centering
\includegraphics[width=0.95\textwidth]{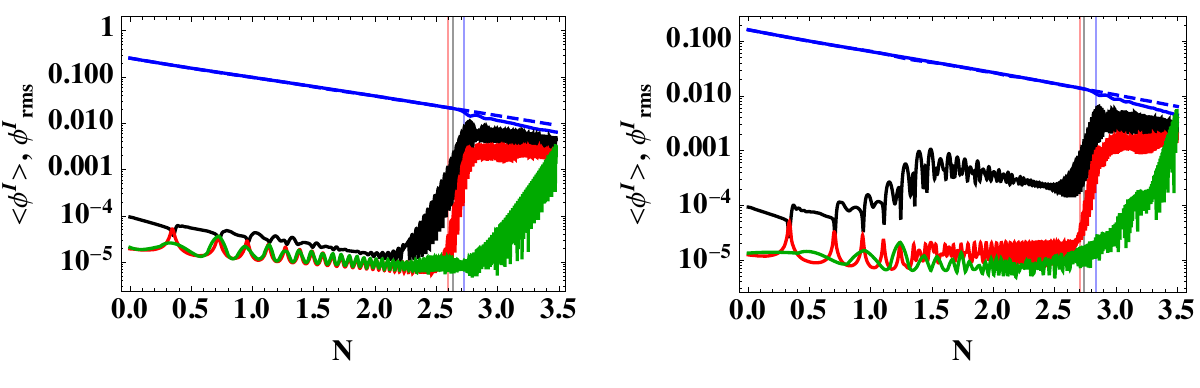}
\caption{\small \baselineskip 11pt
The lattice-averaged background field $\langle \phi \rangle$ (blue), the background field in the linearized analysis $\varphi$ (dashed blue), $\chi_{\rm rms}$ from the lattice (black), $\phi_{\rm rms}$ from the lattice (red) and $\phi_{\rm rms}$ from the linearized analysis (green) for Case A, versus $e$-folds $N$ after the end of inflation. The values of the nonminimal couplings are $\xi_\phi=10$ (left) and $\xi_\phi = 25$ (right). The vertical blue, black, and red lines indicate $\tau_{{\rm bg},0.9}$, $\tau_{\chi_{\rm rms},0.5}$ and $\tau_{\phi_{\rm rms},5}$, respectively. }
\label{fig:Aintermediate}
\end{figure}

Interesting effects arising from the competition between the dominant terms in the effective mass of $\chi$ fluctuations are found in the case of $\xi_\phi \sim 10$, which falls in the intermediate-coupling regime \cite{DeCross:2015uza}. In this regime, the two main contributions to the effective mass-squared of the $\chi$ fluctuations, the contribution $m_{\chi,1}^2$ from the potential and the contribution $m_{\chi,2}^2$ from the field-space (Riemann) curvature, are similar in size at the onset of reheating. Since $m_{1,\chi}^2$ and $m_{2,\chi}^2$ oscillate out of phase, preheating is initially suppressed. This effect was identified in Ref.~\cite{DeCross:2016cbs} (in a linearized analysis) and appears in Fig.~\ref{fig:Aintermediate} as the initial behavior of $\chi_{\rm rms}$ for $N \lesssim 2$ $e$-folds after the end of inflation: no resonant growth occurs. Since $m_{\chi,1}^2$ grows faster than $H^2$, whereas $m_{\chi,2}^2 \propto H^2$ \cite{DeCross:2016cbs}, $m_{1,\chi}^2$ comes to dominate after the so-called cross-over time, which was computed in Ref.~\cite{DeCross:2016cbs}. The relation $m_{\chi,1}^2 \gg m_{\chi,2}^2 $ leads to a period of normal parametric resonance, which is shown in Fig.~\ref{fig:Aintermediate}. The case $\xi_\phi =1$ exhibited a resonance for small $k$, but for $\xi_\phi=10$ the resonant growth of $\chi_k$ occurs for values of the comoving wavenumber up to $k \lesssim 20 \, H_\text{end}$. 

We can see in Fig.~\ref{fig:Aintermediate} that $\phi_\text{rms}$ starts to grow (and diverge from the linearized analysis) around $2.5$ $e$-folds, when $\langle \chi\rangle ={\cal O}(10^{-3})M_{\rm pl}$, similar to the case of $\xi_\phi=1$. At the end of the simulation the RMS values of the two fields are comparable. The energy density in the inflaton condensate $\langle \phi\rangle$ at the end of our simulation (approximately $3$ $e$-folds after the end of inflation) still comprises around $30\%$ of the total energy density of the universe, hence preheating in this case does not complete. 

The case $\xi_\phi = 25$ deserves special attention, since it displays two kinds of resonant behavior. Initially, the Riemann contribution to the effective mass, $m_{\chi,2}^2$, is dominant. This leads to amplification of $\chi$ fluctuations every time that the background field crosses the origin ($\varphi=0$), where the field-space curvature exhibits a large local increase (the ``Riemann spike''). This resonant behavior was previously identified in the large-coupling regime and studied extensively (in a linearized analysis) in Refs.~\cite{DeCross:2015uza,DeCross:2016fdz,DeCross:2016cbs}. 

Fig.~\ref{fig:Aintermediate} indeed shows an increase of $\chi_\text{rms}$ for $N\lesssim 1.5$ $e$-folds after the end of inflation. At 1.5 e-folds after the end of inflation, the growth in $\chi_\text{rms}$ shuts off. This is a result of the decrease of $m_{\chi,2}^2$ compared to $m_{\chi,1}^2$, causing destructive interference, as described above for the $\xi_\phi=10$ case. Up to this time, no significant amount of energy has been transferred to the $\chi$ fluctuations yet. Backreaction is not yet effective, so the linearized approximations for $\chi_\text{rms}$ and $\langle \phi \rangle$ remain valid. At 2.5 $e$-folds after the end of inflation, $m_{\chi,2}^2$ has redshifted enough to render destructive interference between the potential and Riemann terms ineffective, and a period of ordinary parametric resonance starts. This again leads to growth of $\chi_\text{rms}$, which quickly drives growth of $\phi_\text{rms}$ from rescattering.

\subsubsection{Large-coupling regime}

The value $\xi_\phi=100$ places the system in the large-coupling regime \cite{DeCross:2015uza}, which is characterized by a large amplification of $\chi$ fluctuations. Preheating concludes after a few oscillations (5-6 zero-crossings) of the background field, as shown in Figs.~\ref{fig:illustratetimescalesA} and \ref{fig:Arho1100}, and backreaction becomes significant even more quickly.

As validated by the results plotted in Fig.~\ref{fig:illustratetimescalesA}, the background inflaton field $\langle \phi\rangle$ starts to be affected by backreaction effects after the first two oscillations, and becomes completely subdominant shortly thereafter. The variances of the two fields $\phi$ and $\chi$ are almost identical at late times, signaling a kind of ``equipartition'' phenomenon, which can be attributed to the tendency of the system to thermalize. The energy density remaining in the background field is several orders of magnitude smaller than the total energy density of the system, hence we can safely conclude --- within the limitations of the resolution of our lattice --- that the Universe completely preheats into an equilibrated ensemble of $\chi$ and $\phi$ particles for $\xi_\phi=100$. Furthermore, the lattice-averaged value of the $\chi$ field, $\langle \chi \rangle$, which plays the role of the background $\chi$ field in a linearized analysis, remains subdominant to $\langle \phi\rangle$ until the end of the simulation, at which point both are vastly subdominant to the corresponding variances: $\langle \chi\rangle < \langle \phi\rangle\ll \phi_{\rm rms} \simeq \chi_{\rm rms}$. 

\subsubsection{Preheating, Rescattering \& Observables}

The usual comparison of predictions for primordial observables from inflationary models and high-precision observations hinges on the freeze-out of the adiabatic fluctuations after they exit the Hubble radius during inflation. However, this can be violated in multifield models of inflation, due to interactions with isocurvature modes. The interplay between isocurvature and adiabatic modes has been studied extensively in the literature, mostly focusing on the inflationary regime. (For reviews, see Refs.~\cite{Wands:2007bd,Gong:2016qmq}.) In the class of multifield models we study here, the generic single-field attractor behavior strongly suppresses the amplification of isocurvature modes during inflation and the generation of significant non-Gaussianities; these multifield effects occur only for highly fine-tuned choices of couplings and initial conditions \cite{Kaiser:2012ak,Kaiser:2013sna,Schutz:2013fua}.

The adiabatic and entropic (isocurvature) modes are described by the gauge-invariant curvature perturbation $\mathcal{R}_c$ and the normalized entropy perturbation $\mathcal{S}$. On super-horizon scales, the evolution of these petrurbations can be approximated as
\beq
{d{\mathcal{R}_c}\over dN} \simeq 
{2\omega \over H} {\mathcal{S}}
\, , \quad 
{d{\mathcal{S}}\over dN} \simeq 
{\beta} {\mathcal{S}}\, ,
\label{eq:Rc}
\eeq
where $\omega$ is the (covariant) turn-rate of the background trajectory and $\beta$ depends on the slow-roll parameter $\epsilon$, the effective masses of the adiabatic and isocurvature fluctuations, and the turn-rate $\omega$ \cite{Kaiser:2012ak}. In the case of preheating, super-horizon isocurvature modes (with $k\to 0$) can be excited, due to the super-horizon correlation length of the inflaton condensate \cite{Bassett:1998wg,Bassett:1999mt,Bassett:1999ta,Gordon:2000hv}. If the background trajectory is turning, power from the isocurvature modes can be imprinted on the adiabatic modes, following Eq.~\eqref{eq:Rc}. Since we observe significant growth of isocurvature fluctuations after the end of inflation, across a large range of scales $k$, it is imperative to thoroughly check whether predictions for CMB observables are protected or spoiled by interactions during the preheating phase. 

In order to assess the background motion during preheating and identify possible deviations from the single-field attractor, we compute the covariant turn-rate for the system in field-space and consider the ratio $\vert \omega \vert / H$ as we vary $\xi_\phi$. If $\vert \omega \vert /H \ll 1$, the transfer of energy from the isocurvature to the adiabatic modes remains suppressed and thus, at least in the regime of validity of Eq.~\eqref{eq:Rc}, the adiabatic modes on scales relevant to CMB observations do indeed remain frozen during preheating.

In the typical covariant approach for multifield models (which we followed in Refs.~\cite{Kaiser:2012ak,Greenwood:2012aj,Kaiser:2013sna,Schutz:2013fua}), when working to linear order in fluctuations, one defines the length of the background-field velocity vector
\beq
\vert \dot{\varphi}^I \vert \equiv \dot{\sigma} = \sqrt{ {\cal G}_{IJ} \, \dot{\varphi}^I \, \dot{\varphi}^J } \, ,
\label{dotsigmadef}
\eeq
in terms of which one may define the unit vector
\beq
\hat{\sigma}^I \equiv \frac{ \dot{\varphi}^I }{\dot{\sigma}} .
\label{hatsigma}
\eeq
The covariant turn-rate vector is then given by
\beq
\omega^I \equiv {\cal D}_t \hat{\sigma}^I = - \frac{1}{ \dot{\sigma}} V_{, K} \, \hat{s}^{IK} \, ,
\label{omegadef}
\eeq
where the last expression follows upon using the equation of motion for the homogeneous background fields $\varphi^I$, and the quantity $\hat{s}^{IK} \equiv {\cal G}^{IK} - \hat{\sigma}^I \hat{\sigma}^K$ projects into directions of field-space orthogonal to $\hat{\sigma}^I$. 
Using the properties of the projectors $\hat{s}^{IK}$ \cite{Kaiser:2012ak}, it is straightforward to show that
\beq
\vert \omega^I \vert = \frac{1}{ \dot{\sigma}^2} \left[ V_{, K} \, V_{, L} \left( \dot{\sigma}^2 {\cal G}^{KL} - \dot{\varphi}^K \, \dot{\varphi}^L \right) \right]^{1/2} \, .
\label{turnrate2}
\eeq
The only term in Eq.~(\ref{turnrate2}) that becomes problematic during preheating is the coefficient $1/\dot{\sigma}^2$, outside the square brackets, since during preheating $\dot{\sigma}$ repeatedly oscillates to zero. However, if we perform a time-average of this coefficient over each oscillation of the inflaton condensate, then the expression for $\vert \omega \vert = \vert \omega^I \vert$ in Eq.~(\ref{turnrate2}) remains well-behaved. To apply Eq.~(\ref{turnrate2}) to our lattice simulations, we evaluate all terms that depend on the spatially homogeneous quantities $(\varphi^I, \dot{\varphi}^I)$ by substituting the corresponding spatial averages on the lattice. 

\begin{figure}
\centering
\includegraphics[width=0.95\textwidth]{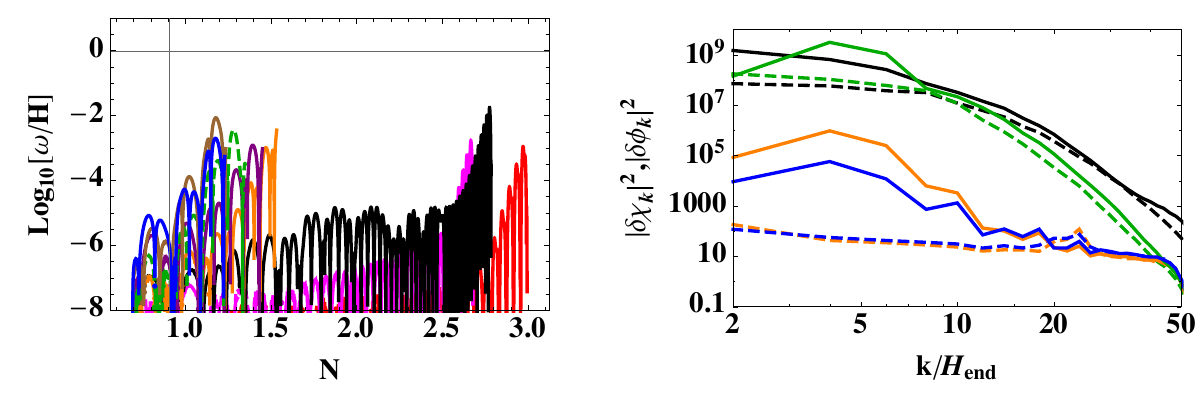}
\caption{\small \baselineskip 11pt
{\it Left:} The covariant turn-rate $\vert \omega \vert$ normalized by the Hubble scale $H$ for Case A and different values of the nonminimal coupling in the range $1\le\xi_\phi\le 100$. Color coding as in Fig.~\ref{fig:Arho1100}. We see that in all cases $\vert \omega \vert < 0.1 H$. Note that, as explained in the main text, we only consider the turn-rate for times such that $\langle \phi^I\rangle > \phi^I_{\rm rms}$ for at least one component $I$, since this defines the regime of validity of Eqs.~\eqref{eq:Rc} and \eqref{turnrate2}. {\it Right:} The power spectrum for $\phi$ (dashed) and $\chi$ (solid) fluctuations for $\xi_\phi=10$ and $N= 2.3, 2.4, 2.7, 3.0$ (blue, orange, green, black, respectively). We see that when backreaction becomes significant the two spectra become nearly indistinguishable.}
\label{fig:AomegaoverH}
\end{figure}

Eqs.~(\ref{eq:Rc}) and (\ref{turnrate2}) only apply when one may identify a well-defined background solution. In the context of our lattice simulations, Eqs.~(\ref{eq:Rc}) and (\ref{turnrate2}) remain well-defined for times such that the lattice-averaged value for at least one component of $\phi^I = \{ \phi, \chi \}$ remains greater than the corresponding RMS value for that component: $\langle \phi^I \rangle > \phi_{\rm rms}^I$ for at least one component $I$. (We consider this condition further in Section~\ref{sec:parameters} when we examine Case D, with positive ellipticity.) Subject to this criterion, we find the results for $\omega$ (normalized by the Hubble rate) shown in the left panel of Fig.~\ref{fig:AomegaoverH}. We see that for all values of $\xi_\phi$ under consideration, the turn-rate remains much smaller than the Hubble scale throughout the preheating phase, even as $H$ itself falls over time.

A second criterion for whether preheating dynamics could affect predictions for primordial observables concerns how quickly the system reaches the adiabatic limit \cite{Elliston:2011dr,Meyers:2013gua,Elliston:2014zea,Turzynski:2014tza}. The right panel of Fig.~\ref{fig:AomegaoverH} shows that soon after the $\chi$ particles are produced, they rescatter efficiently, producing nearly identical spectra for the $\delta \chi_k$ and $\delta \phi_k$ modes. (This is consistent with our finding in Fig.~\ref{fig:Aintermediate} that $\phi_{\rm rms} \simeq \chi_{\rm rms}$ within a few $e$-folds after the end of inflation for $\xi_I \sim {\cal O} (10)$.) Although an accurate simulation of thermalization is beyond the scope of this work, Fig.~\ref{fig:AomegaoverH} indicates that an ``equipartition" of energy density quickly emerges between the two fields, which can be considered a precursor to thermalization \cite{Micha:2002ey,Micha:2004bv,McDonough:2020tqq}. Within the first few $e$-folds after the end of inflation, the fluctuations in both the $\chi$ and $\phi$ fields lose all information about the wavenumber-dependent structure of the initial resonances, and begin to approach quasi-thermal spectra. Since the turn-rate remains tiny up through the time that quasi-thermal spectra emerge, we conclude that the predictions for observables such as the spectral index $n_s$ and the tensor-to-scalar ratio $r$ from models of the type in Case A, across a wide range of nonminimal couplings $\xi_\phi$, remain unaffected by the rapid growth of isocurvature fluctuations after the end of inflation.

In sum, we have shown that for the ``generic'' benchmark case (Case A), preheating becomes more efficient as we increase $\xi_\phi$. Different dynamics occur for different regimes of $\xi_\phi$, governed by the interplay between contributions to $m_{{ \rm eff}, I}^2 (t)$ from the potential and the field-space structure. Several nonlinear processes unfold over distinct time-scales, some of which are strongly correlated with each other.  For large values of the nonminimal coupling, $\xi_\phi \sim 100$, we find that each of these processes is essentially completed within $N_{\rm reh} \lesssim 2.5$ $e$-folds after the end of inflation.

\subsection{Parameter Dependence}
\label{sec:parameters}

Having presented a detailed analysis of the preheating dynamics for the ``benchmark'' case (Case A in Table~\ref{tab:potentialtopography}), we now examine how preheating depends on the potential topography. Ref.~\cite{DeCross:2016fdz} analyzed parametric resonance for this family of models to linear order in fluctuations, and identified distinct characteristic behavior across parameter space. Building on the intuition afforded by that analysis, we now examine the fully nonlinear dynamics of each of the four characteristic cases (B-D) shown in Table~\ref{tab:potentialtopography}, and compare them with the benchmark Case A.

\begin{figure}
\centering
\includegraphics[width=0.95\textwidth]{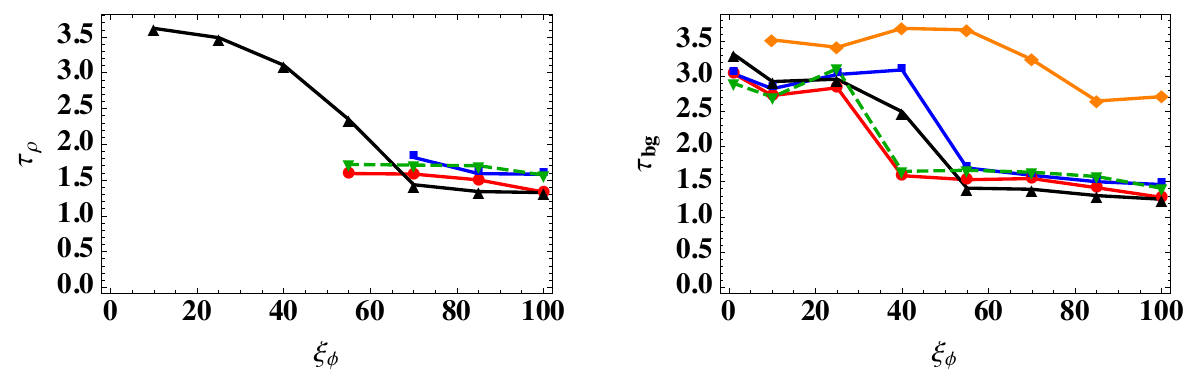}
\caption{\small \baselineskip 11pt
$\tau_{\rho,0.95}$ (left) and $\tau_{\text{bg},0.9}$ (right), for all five cases. Color coding is Case A (benchmark case, red), Case B (symmetric couplings, blue), Case C (negative ellipticity, orange), Case D (positive ellipticity, black) and Case E (zero ellipticity, dashed green).}
\label{fig:AEtimescales}
\end{figure}

Fig.~\ref{fig:AEtimescales} shows the dependence of the time-scales $\tau_\rho$ and $\tau_\text{bg}$ on the value of the nonminimal coupling $\xi_\phi$ for each of the five characteristic cases (A-E). We see that complete energy transfer is only possible for $\xi_\phi \gtrsim 55$, except for Case D (positive ellipticity), which is particularly efficient. Preheating in Case C (negative ellipticity), on the other hand, proceeds slowly, and complete energy transfer does not occur even for large $\xi_\phi$. The backreaction of produced particles on the evolution of the inflaton condensate is also significantly slower for Case C compared to the other cases.
Despite differences among the cases, however, we again find the general trend that preheating becomes more efficient as the values of the nonminimal couplings $\xi_I$ increase, and that (apart from Case C with negative ellipcticity) the relevant, nonlinear preheating processes are completed within $N_{\rm reh} \lesssim 2$ in the limit $\xi_\phi \sim 100$.

\begin{figure}
    \centering
    \includegraphics[width = 0.95 \textwidth]{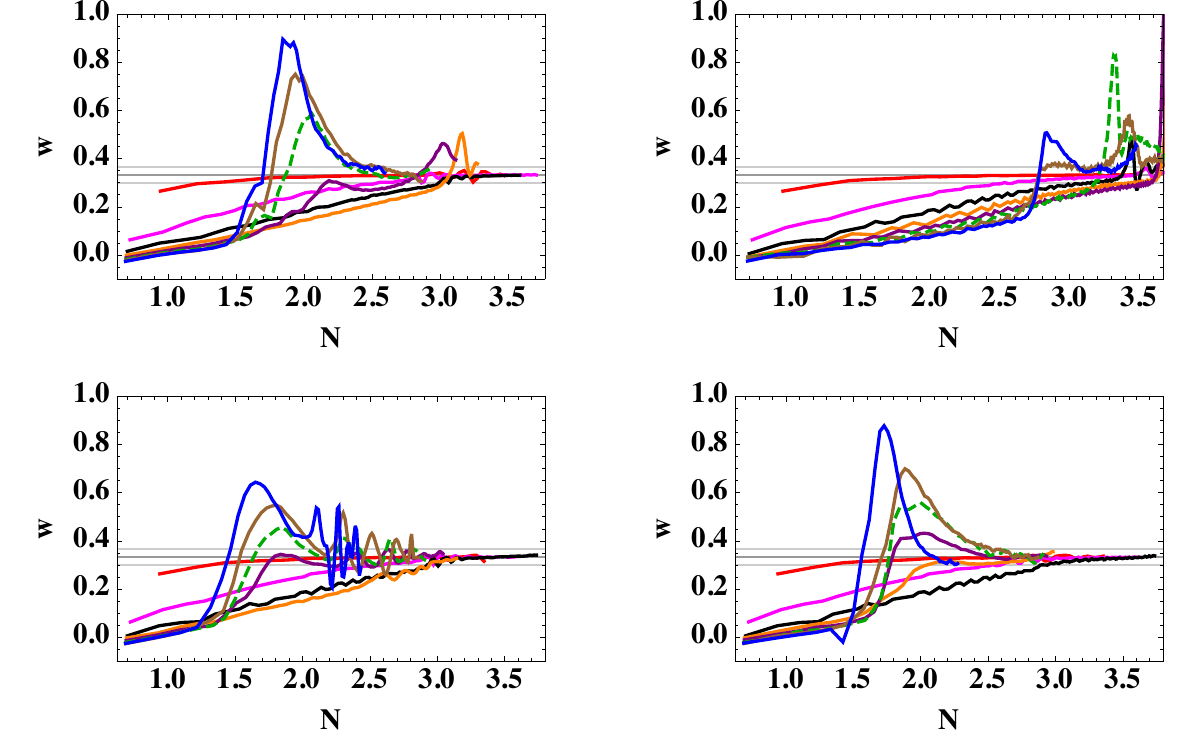}
    \caption{\small \baselineskip 11pt Equation of state for Case B (symmetric couplings, top left), Case C (negative ellipticity, top right), Case D (positive ellipticity, bottom left) and Case E (zero ellipticity, bottom right). Color coding as in Fig.~\ref{fig:Arho1100}. The averaging window is always the period of the $\phi$-field. For Case D, the lattice-averaged value of the $\chi$-field becomes larger than the $\phi$-field for some values of $\xi_\phi$. The rapid oscillations of $w$ for $\xi_\phi = 85, 100$ can be removed by switching to an integration window corresponding to the oscillations of $\langle \chi \rangle$ once $\langle \chi \rangle$ dominates over $\langle \phi \rangle$.}
    \label{fig:4eos}
\end{figure}

Fig.~\ref{fig:4eos} shows the evolution of the equation of state, averaged over each oscillation of the $\phi$ field. The approach of the equation of state to radiation-dominated expansion ($w = 1/3$) proceeds similarly to the benchmark Case A across these cases (compare with Fig.~\ref{fig:Arho1100}, right panel), except for Case C (negative ellipticity). We explore each of these cases in more detail below.

\subsubsection{Case B: Symmetric Couplings}

A physically important case is that of symmetric couplings ($\xi_\phi = \xi_\chi$ and $\lambda_\phi = g = \lambda_\chi$), which arises if the  model obeys an $SO(2)$ (or higher) symmetry. This would be a proxy for Higgs inflation, if one focused on the scalar degrees of freedom and neglected the effects of the gauge bosons \cite{DeCross:2016cbs}. (See Ref.~\cite{Sfakianakis:2018lzf} for a detailed computation of preheating in Higgs inflation; compare with Refs.~\cite{Bezrukov:2008ut,GarciaBellido:2008ab,Dufaux:2010cf,Repond:2016sol}.) As shown in Ref.~\cite{Nguyen:2019kbm}, the symmetric case (Case B) has a similar preheating behavior to the benchmark case (Case A) for $\xi_\phi=1,10,100$, although preheating occurs with a minor delay in the symmetric case, with a slightly longer time-scale before significant mode amplification dominates the dynamics. 

\begin{figure}
    \centering
    \includegraphics[width = 0.95\textwidth]{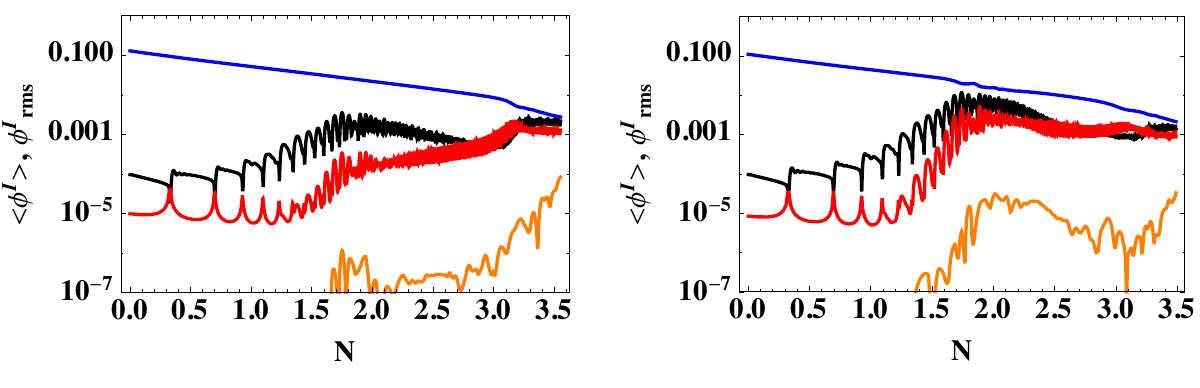}
    \caption{\small \baselineskip 11pt The lattice-averaged fields and corresponding RMS values versus $e$-folds after the end of inflation for Case B (symmetric couplings), with $\xi_\phi=40,55$ (left and right, respectively). We have used interpolation between the peaks of $\langle\phi\rangle$ (blue), $\langle\chi\rangle$ (orange) and $\phi_\text{rms}$ (red), $\chi_\text{rms}$ (black).}
    \label{fig:Bvariances}
\end{figure}

Figure~\ref{fig:4eos} shows the averaged equation of state for the symmetric Case B. We see a similar trend to the benchmark Case A: a temporarily stiff equation of state for large values of $\xi_\phi$, which quickly asymptotes to a radiation-dominated one with $w\simeq 1/3$. However, for $\xi_\phi=40$, there emerges a late increase in $w$ after $N=3$ $e$-folds, whereas for $\xi_\phi=55$ the equation of state exhibits a temporary local maximum at $N\simeq 2$ and then exceeds $w=1/3$ at $N\simeq 3$. This behavior can be easily understood by considering the field variances for these values of $\xi_\phi$, as shown in Fig.~\ref{fig:Bvariances}. 

For both $\xi_\phi=40, 55$ we see an early period of amplification of the $\chi$ field driven by the field-space curvature (Riemann) term $m_{2,\chi}$ in the effective mass. This resonance is too weak to completely preheat the universe and eventually shuts off. At later times, after the Riemann contribution has redshifted to become a subdominant contribution to $m_{\rm eff, \chi}$, a period of ordinary (potential-driven) parametric resonance drives a significant transfer of energy from the inflaton into higher-momentum $\chi$ modes. This causes the late-time peak in the equation of state. In the case $\xi_\phi=55$ the initial resonance at $N\simeq 1.5$ is strong enough to drain sufficient energy from the inflaton condensate and drive a local peak in the equation of state. It is interesting to note that the late-time parametric resonance occurs in the $\phi$ field. This two-stage parametric resonance process is similar to that found above for the benchmark Case A for $\xi_\phi \sim 25$ (see Section~\ref{sec:intermediate}).

\subsubsection{Case C: Negative Ellipticity}

The negative ellipticity case (with $\varepsilon = (\xi_\phi - \xi_\chi) / \xi_\phi < 0$) shows the largest deviation from the benchmark Case A, as well as the other cases that we explore here. This can be seen in the time-scales shown in Fig.~\ref{fig:AEtimescales} as well as the equation of state shown in Fig.~\ref{fig:4eos}. These results are consistent with our previous findings from our linearized analysis \cite{DeCross:2016fdz}, in which we had found that a large negative ellipticity suppresses the Floquet structure of the model, almost eliminating all resonance bands. This is verified by the fully nonlinear lattice simulations, in which we find that complete preheating is not reached, even for $\xi_\phi=100$, contrary to all other cases. Furthermore, the averaged equation of state remains $w<1/3$ through $N\approx 2.5$ $e$-folds after the end of inflation. In this case, we find a modest, late-time parametric resonance, after the Rieman spike has redshifted enough to be insignificant. 

\begin{figure}
    \centering
    \includegraphics[width = 0.95\textwidth]{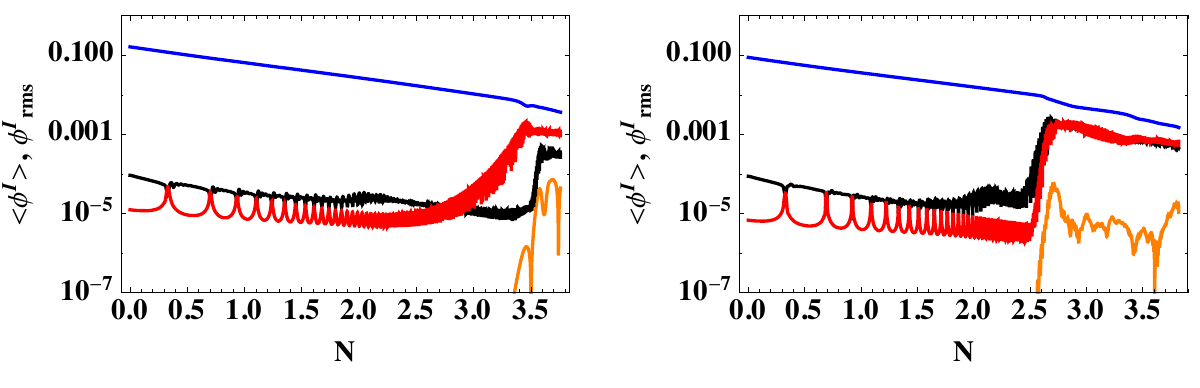}
    \caption{\small \baselineskip 11pt The lattice-averaged fields and corresponding RMS values versus $e$-folds after the end of inflation for Case C (negative ellipticity), with $\xi_\phi=25,85$ (left and right, respectively). We have used interpolation between the peaks of $\langle\phi\rangle$ (blue), $\langle\chi\rangle$ (orange) and $\phi_\text{rms}$ (red), $\chi_\text{rms}$ (black).}
    \label{fig:Cvariances}
\end{figure}

It is interesting to note that for small values of $\xi_\phi \lesssim 25$, the late-time resonance occurs in the $\phi$ field, while for larger values of $\xi_\phi$ the late-time resonance is driven primarily by the $\chi$ field; in the latter case, the resonance is driven predominantly by the potential contribution to the effective mass, $m_{1, \chi}$, rather than from the field-space curvature contribution, $m_{2, \chi}$. Fig.~\ref{fig:Cvariances} shows the behavior of the spatially averaged fields and their corresponding RMS values for two characteristic values of $\xi_\phi$ to illustrate these two different regimes.

\subsubsection{Case D: Positive Ellipticity}

We considered the case of positive ellipticity within a linearized analysis in Ref.~\cite{DeCross:2016fdz}, and found that positive ellipticity led to stronger violation of the adiabaticity condition and, correspondingly, larger resonance bands than models with symmetric couplings, yielding efficient preheating. We find comparable behavior in our nonlinear lattice simulations as well. For example, Fig.~\ref{fig:AEtimescales} shows that $\tau_{\rho}$ occurs more quickly for Case D than for any of the other cases in the regime of large couplings $\xi_\phi \geq 55$. Even more striking, only for Case D does preheating lead to a significant transfer of energy density from the inflaton condensate into radiative degrees of freedom for smaller nonminimal couplings, $\xi_\phi \sim {\cal O} (10)$.

\begin{figure}
    \centering
    \includegraphics[width = 0.95\textwidth]{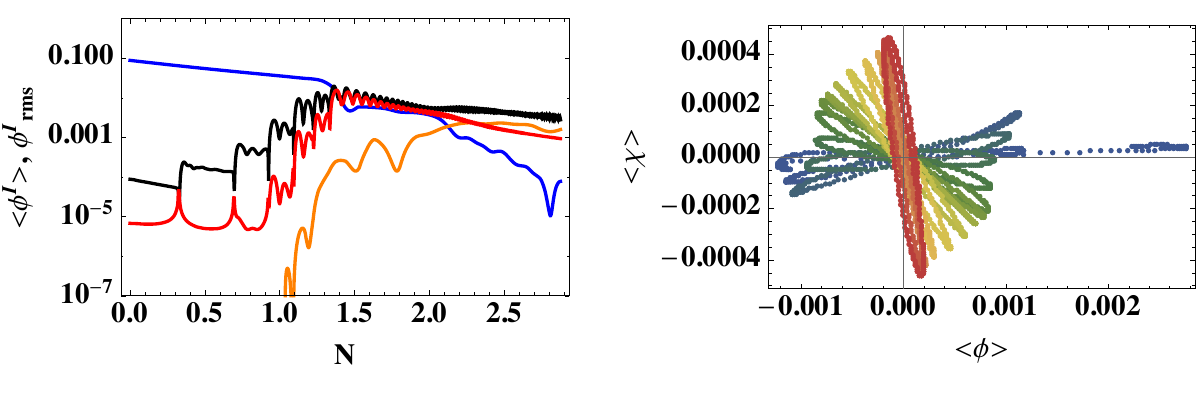}
    \caption{\small \baselineskip 11pt  {\it Left:}  The lattice-averaged fields and corresponding RMS values versus $e$-folds after the end of inflation for Case D (positive ellipticity) with  $\xi_\phi=85$. We have used interpolation between the peaks of $\langle\phi\rangle$ (blue), $\langle\chi\rangle$ (orange), $\phi_\text{rms}$ (red), and $\chi_\text{rms}$ (black).
    {\it Right:} The lattice-averaged motion in the $\phi-\chi$ plane between 1.3 and 2.7 $e$-folds after the end of inflation. Later times correspond to redder colors.  }
    \label{fig:Dvariances}
\end{figure}

As shown in Fig.~\ref{fig:4eos}, the equation of state $w$ for Case D shows large, persistent oscillations around $w\simeq 1/3$ after the first $1.5$ $e$-folds. As noted above, the equation of state shown in Fig.~\ref{fig:4eos} in each case was averaged over an oscillation period of the inflaton condensate, $\langle \phi \rangle$. To understand the large oscillations in $w$ for Case D, it suffices to consider the dynamics of this model with $\xi_\phi = 85$. The left panel of Fig.~\ref{fig:Dvariances} shows the lattice-average field values $\langle \phi \rangle$ and $\langle \chi \rangle$ as well as the corresponding variances $\phi_{\rm rms}$ and $\chi_{\rm rms}$. We see that after a significant energy transfer has occurred from the inflaton condensate into radiative modes, the lattice-averaged (background) trajectory of the system is primarily aligned along the $\chi$ axis. After this point, the dominant frequency for the system is that of $\langle \chi \rangle$ oscillations, rather than the $\langle \phi \rangle$ oscillations. If we average the equation of state $w$ within a window set by the oscillation frequency of $\langle \chi \rangle$, we find that $w$ 
approaches $w = 1/3$ in a way more comparable to that of the benchmark Case A.

The apparent departure from the single-field attractor along $\langle \chi\rangle \simeq 0$ raises the question of whether predictions for CMB observables in this case might be altered by post-inflationary dynamics. Even for Case D, however, the strong agreement between predictions from this family of models and CMB observations remains protected, despite the violent energy exchange after the end of inflation, for at least two reasons.

First, we note that the power spectra of the $\phi$ and $\chi$ fields have ``equlibrated'' (akin to the spectra shown in Fig.~\ref{fig:AomegaoverH}) prior to the time when $\langle \chi \rangle$ dominates over $\langle \phi\rangle$, hence no information can be transmitted coherently from subhorizon to superhorizon scales. (In Fig.~\ref{fig:Dvariances} this effect is consistent with $\phi_{\rm rms} \simeq \chi_{\rm rms}$ by $N \simeq 1.3$ $e$-folds after inflation.) Although a detailed discussion of equilibration and thermalization is beyond the scope of the present work, it is clear from such spectra that the final distribution of power among the $\phi$ and $\chi$ fluctuations is vastly different from that following the initial parametric resonance. Strong rescattering effects appear to have largely erased any wavenumber-dependent information and yielded spectra that are similar among the $\phi$ and $\chi$ fields.

\begin{figure}
    \centering
    \includegraphics[width = 0.55\textwidth]{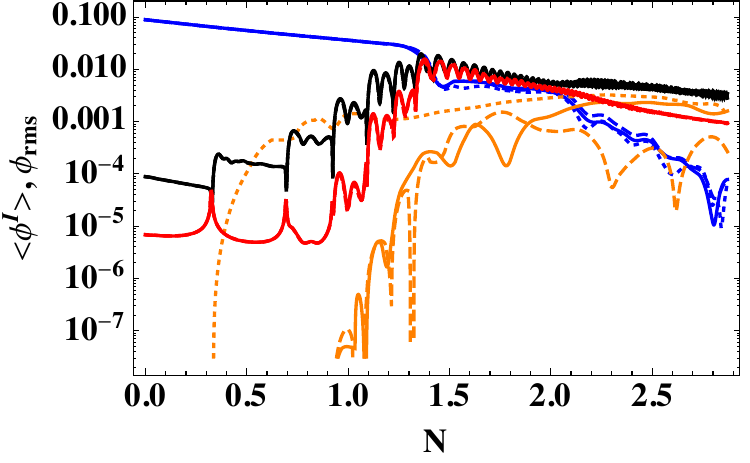}
    \caption{\small \baselineskip 11pt Lattice-averaged field values and corresponding RMS values for Case D (positive ellipticity) with $\xi = 85$ for three different seeds for the initial fluctuations. Color coding is the same as in the left panel of Fig.~\ref{fig:Dvariances}. The three different seeds are shown in solid, dashed and dotted. Note that across the three seeds, the late-time values of $\phi_{\rm rms}$ (red) and $\chi_{\rm rms}$ (black) remain consistent, whereas the late-time values of $\langle \chi \rangle$ (orange) differ significantly, indicating a lack of superhorizon coherence of the $\chi$ field after the end of inflation.}
    \label{fig:D3seeds}
\end{figure}

Second, we have found that in all cases, $\chi_{\rm rms} \gg \langle \chi\rangle$ by the time $\chi_{\rm rms} \simeq \phi_{\rm rms}$. This implies that the evolution of the lattice-averaged quantity $\langle \chi \rangle$ is a statistical phenomenon that assumes different values in each Hubble patch, rather than a coherent motion of the $\chi$ field across superhorizon scales. In order to examine this, we performed simulations with different seeds for the fluctuations, all taken from the same distribution. Fig.~\ref{fig:D3seeds} shows the results for three such simulations. We see that while the variances are identical, the lattice-averaged values depend strongly on the specific seed. Hence different Hubble patches, which in our simulation can be simulated by selecting different seeds, will develop different values of $\langle \chi\rangle$. This suggests that no $\chi$ condensate exists with superhorizon correlations, which in turn implies that the power in low-$k$ modes cannot be imprinted onto the adiabatic perturbations on CMB-relevant scales after the end of inflation. 

More fundamentally, while our analysis shows that predictions for CMB observables in this family of models remain unaffected by preheating, we emphasize that the formulas in Eq.~(\ref{eq:Rc}) that describe the superhorizon evolution of adiabatic and isocurvature perturbations presuppose the existence of a well-defined, coherent background motion, in terms of which one may define the covariant turn-rate $\omega$, as in Eq.~(\ref{turnrate2}). When the RMS of the fields far exceed the lattice-averaged field values, the existence of a coherent background motion becomes ill-defined. How to modify simple relations like Eq.~(\ref{eq:Rc}) for evolution deep into the nonlinear regime remains an interesting question, beyond the scope of the present work.

\subsubsection{Case E: Zero Ellipticity}

We conclude our analysis with Case E (zero ellipticity), for which $\xi_\phi=\xi_\chi$, but for which, in general, $\lambda_\phi \neq g \neq \lambda_\chi$ (see Table \ref{tab:potentialtopography}). Because the couplings in the Jordan-frame potential do not equal each other, Case E is distinct from the symmetric Case B. Nonetheless, we see that the phenomenology for Case E is largely indistinguishable from the benchmark Case A, both in terms of the relevant time-scales (Fig.~\ref{fig:AEtimescales}) as well as the evolution of the equation of state (Figs.~\ref{fig:Arho1100} \& \ref{fig:4eos}). This result is not surprising, since any non-adiabatic behavior caused by the Riemann spike in $m_{2, \chi}$ is not very sensitive to small changes among the couplings in the Jordan-frame potential. The Riemann spike, after all, derives from the curvature of the field-space manifold, which depends only on the nonminimal couplings $\xi_I$.

\section{Conclusions}
\label{sec:conclusions}

Post-inflation reheating is a critical phase in the early history of our universe, connecting the era of primordial inflation with standard Big Bang evolution. The energy density that had driven accelerated expansion during inflation must be dispersed rapidly into a hot, thermal bath of Standard Model (and perhaps Dark Matter) particles, and the equation of state must assume a radiation-dominated form. Any significant delay between the end of inflation and the conclusion of reheating would have implications for the comparison of predictions for observables and high-precision measurements, by altering the expansion history of the early universe \cite{Bassett:2005xm,Allahverdi:2010xz,Frolov:2010sz,Amin:2014eta,Lozanov:2019jxc}. Yet the principal physical processes during the reheating phase --- energy transfer from the inflaton condensate into higher-momentum particles, emergence of a radiation-dominated equation of state, and thermalization of the produced particles at an appropriately high temperature --- can unfold according to their own distinct time-scales, which depend in different ways on model parameters. 

In this work we have examined the post-inflation phase in a family of models that incorporate realistic features from high-energy theory: multiple interacting scalar degrees of freedom, and nonminimal couplings between those fields and the space-time curvature. We have investigated several distinct nonlinear processes in these models using lattice simulations, and explored how the relevant time-scales vary across parameter space. Across all the variations within this family of models, we consistently find efficient preheating in the limit of large nonminimal couplings $\xi_I \sim {\cal O} (100)$, with each of the requisite preheating effects (energy transfer, radiation-dominated equation of state, and onset of thermalization) completed within $N_{\rm reh} \lesssim 3$ $e$-folds after the end of inflation. Moreover, across the variations in models and regions of parameter space, inflationary predictions for observable features in the cosmic microwave background radiation remain unaffected by the violent post-inflationary dynamics, thereby protecting the close match \cite{Kaiser:2013sna} between these models and the latest observations.

In general, the post-inflationary behavior of most models within this family behave in comparable ways to the ``benchmark" or generic case on which we first focused (Case A) \cite{Nguyen:2019kbm}. We found significant deviations from the benchmark case only for large, negative ellipticity (Case C), which can arise if there is a misalignment between the dominant direction in field-space along which the system evolves and the larger of the nonminimal couplings, $\xi_I$. In such cases, the parametric resonances that are typical of preheating become quenched. We further identified more modest differences from the benchmark case for large, positive ellipticity (Case D), in which particle production was even more efficient than average. In that case, the superhorizon coherence of the inflaton condensate was lost even more quickly than in more ``generic" cases.

A physical effect that we have not considered in this work, but which could have interesting consequences for preheating in particular regions of parameter space, arises from coupled metric perturbations. In Ref.~\cite{DeCross:2016cbs} the detailed effects of the various contributions to the effective masses $m_{\rm eff, \phi}^2$ and $m_{\rm eff,\chi}^2$ were analyzed for this family of models to linear order in the perturbations. The coupled metric perturbations were shown to affect the adiabatic fluctuations  ($\delta\phi$) for low wavenumbers (super-horizon modes, with $k < aH$). On the other hand, in these models the growth of isocurvature modes typically occurs efficiently across a wide range of wavenumbers, and is typically more efficient than the growth of adiabatic modes for large values of the nonminimal couplings $\xi_I \gg 1$. We therefore expect that the effect of coupled metric perturbations may be most pronounced in cases with $\xi_I \gg 1$, in those regions of parameter space in which the growth of isocurvature modes is suppressed due to a negative ellipticity, as in Case C; otherwise we expect effects from metric perturbations to remain subdominant, compared to the nonlinear field dynamics analyzed here. Recently, the limitations of linearized gravity for treating violent processes such as preheating has been discussed, for example in Ref.~\cite{Giblin:2019nuv}.
A full numerical treatment of gravitational effects in preheating in this family of models remains the subject of further research.

Other natural extensions of this work would include coupling the scalar fields to higher-spin fields, including both fermions \cite{Greene:1998nh,Greene:2000ew,Peloso:2000hy,Tsujikawa:2000ik,Adshead:2015kza,Adshead:2017znw,Repond:2016sol} and gauge bosons \cite{Davis:2000zp,GarciaBellido:2003wd,Braden:2010wd,Allahverdi:2011aj,Deskins:2013lfx,Adshead:2015pva,Adshead:2017xll,Cuissa:2018oiw}, and exploring possible observational consequences, such as the amplification of gravitational waves during preheating. (See, e.g., Refs.~\cite{Adshead:2018doq,Lozanov:2019ylm} and references therein.) In addition, although several quite distinct families of ``attractor" models of inflation produce comparable predictions for CMB spectra if one only considers their dynamics during inflation \cite{Christodoulidis:2019jsx,Christodoulidis:2019mkj}, some of these other models (beyond the family we studied here) would likely {\it not} feature such efficient preheating \cite{Iarygina:2020dwe}. Studies like this one may therefore begin to differentiate among competing families of models and clarify how predictions for CMB observables might be affected by distinct reheating scenarios.

\section*{ Acknowledgements}
It is a pleasure to thank Mustafa Amin and Kaloian Lozanov for helpful discussions. JvdV was funded by the Deutsche Forschungsgemeinschaft under Germany's Excellence Strategy - EXC 2121 €œQuantum Universe - 390833306.
EIS acknowledges support from the Dutch Organisation for
Scientific Research (NWO).
A portion of this research was conducted in MIT's Center for Theoretical Physics and supported in part by the U.S. Department of Energy under Contract No.~DE-SC0012567. JTG and RN are supported by the National Science Foundation PHYS-1719652.  JTG would also like to thank MIT and MIT's Center for Theoretical Physics for gracious hospitality while some of this work was completed.

\appendix

\section{Field-Space Metric and Related Quantities}
\label{sec:AppendixAFieldSpace}

Given $f (\phi^I)$ in Eq. (\ref{f2field}) for a two-field model, the field-space metric in the Einstein frame, Eq. (\ref{GIJ}), takes the form
\beq
{\cal G}_{IJ}=
\left( \frac{M_{\rm pl}^2}{4f^2} \right)
\begin{pmatrix} 
2f+6\xi_\phi^2 \phi^2 & 6 \xi_\phi \xi_\chi \phi \chi \\
6 \xi_\phi \xi_\chi \phi \chi & 2f+6\xi_\chi^2 \chi^2
\end{pmatrix} \, ,
\label{Gphiphi}
\eeq
where $\phi^I \equiv \{\phi,\chi\}  $. The inverse metric is
\beq
{\cal G}^{IJ}=
\left( \frac{2f}{M_{\rm pl}^2C} \right)
\begin{pmatrix} 
2f+6\xi_\chi^2 \chi^2 & -6 \xi_\phi \xi_\chi \phi \chi \\
-6 \xi_\phi \xi_\chi \phi \chi & 2f+6\xi_\phi^2 \phi^2
\end{pmatrix} \, ,
\label{inverseG}
\eeq
where $C (\phi^I)$ is defined as
\beq
\begin{split}
C (\phi, \chi) &\equiv M_{\rm pl}^2 + \xi_\phi (1 + 6 \xi_\phi) \phi^2 + \xi_\chi (1 + 6 \xi_\chi ) \chi^2 \\
&= 2f + 6 \xi_\phi^2 \phi^2 + 6 \xi_\chi^2 \chi^2 .
\end{split}
\label{C}
\eeq

The Christoffel symbols for this field space take the form
\beq
\begin{split}
\Gamma^\phi_{\>\> \phi \phi} &= \frac{\xi_\phi (1 + 6 \xi_\phi ) \phi}{C} - \frac{\xi_\phi \phi}{f}\, , \\
\Gamma^\phi_{\>\> \chi \phi} = \Gamma^\phi_{\>\> \phi \chi} &= - \frac{\xi_\chi \chi}{2f}\, , \\
\Gamma^\phi_{\>\> \chi \chi} &= \frac{\xi_\phi (1 + 6 \xi_\chi) \phi}{C} , \\
\Gamma^\chi_{\>\> \phi \phi} &= \frac{\xi_\chi (1 + 6 \xi_\phi) \chi}{C}\, , \\
\Gamma^\chi_{\>\> \phi \chi} = \Gamma^\chi_{\>\> \chi \phi} &= - \frac{\xi_\phi \phi}{2f} \, , \\
\Gamma^\chi_{\>\> \chi \chi} &= \frac{\xi_\chi (1 + 6 \xi_\chi ) \chi}{C} - \frac{\xi_\chi \chi}{f} \, .
\end{split}
\label{Gammas}
\eeq

For two-dimensional manifolds the Riemann tensor can be written in the form
\beq
{\cal R}_{ABCD} = \frac{1}{2} {\cal R} (\phi^I) \left[ {\cal G}_{AC} {\cal G}_{BD} - {\cal G}_{AD} {\cal G}_{BC} \right] ,
\label{Riemann2d}
\eeq
where ${\cal R} (\phi^I)$ is the Ricci scalar. Given the field-space metric of Eq. (\ref{Gphiphi}), we find
\beq
{\cal R} (\phi^I ) = \frac{1}{3 M_{\rm pl}^2 C^2} \left[ (1 + 6 \xi_\phi) (1 + 6 \xi_\chi) (4 f^2 ) - C^2 \right] .
\label{Ricci2d}
\eeq

\section{Adiabaticity Violation induced by Field-Space Curvature}
\label{sec:AppendixBAdiabaticity}

Following the analysis of Ref.~\cite{DeCross:2015uza}, we define the adiabaticity parameter for the $\chi$ fluctuations as
\beq
{\cal A}_\chi \equiv \frac{ \partial_\eta \, \Omega_\chi}{\Omega_\chi^2 } = 
 {H^{-3} \partial_t m_{{\rm eff},\chi}^2 +2 ( m_{{\rm eff},\chi}/H )^2
\over
2 \left [(k/[ a H ])^2 + (m_{{\rm eff},\chi}/H)^2  \right ]^{3/2} } \, ,
\eeq
where $\eta$ is conformal time and the effective frequency $\Omega_\chi (t)$ is given in Eq.~(\ref{OmegaIdef}). For small values of the wavenumber $k\ll aH$ this can be approximated by
\beq
{\cal A}_\chi \simeq {\partial_t m_{{\rm eff},\chi}^2\over 2 m_{{\rm eff},\chi}^3} + {H\over m_{{\rm eff},\chi}} \, .
\eeq
For the parameters considered here, the first term dominates the adiabaticity violation, so we will only consider this, in order to build intuition and connect adiabaticity to the potential topography and field-space geometry. Throughout this section we will use $\tilde \Lambda_\phi=-0.2$.

\begin{figure}[t!]
\centering
\includegraphics[width=0.95\textwidth]{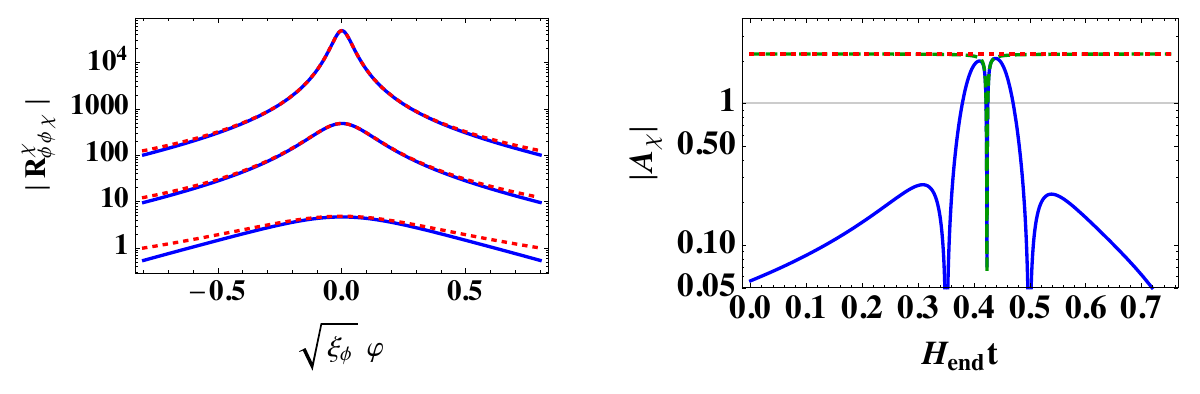}
\caption{ \small \baselineskip 11pt
{\it Left:} The magnitude of the Riemann spike as a function of the rescaled background field $\sqrt{\xi_\phi} \varphi$ for $\xi_\phi=1,10,100$ (bottom to top) and $\xi_\chi=0.8\xi_\phi$. The blue line shows the exact result and the dashed red line the Lorentzian approximation.
{\it Right:} The adiabaticity parameter for $\xi_\phi=100$, $\xi_\chi=80$ around the first inflaton zero-crossing for $k\to 0$. The blue curve shows the full result and the green-dotted shows the result of only considering the Riemann term of the effective mass. The red-dotted line shows the analytic result $2/\sqrt{1-\varepsilon}$.
 }
 \label{fig:Lorentzian}
\end{figure}

\begin{figure}[t!]
\centering
\includegraphics[width=0.95\textwidth]{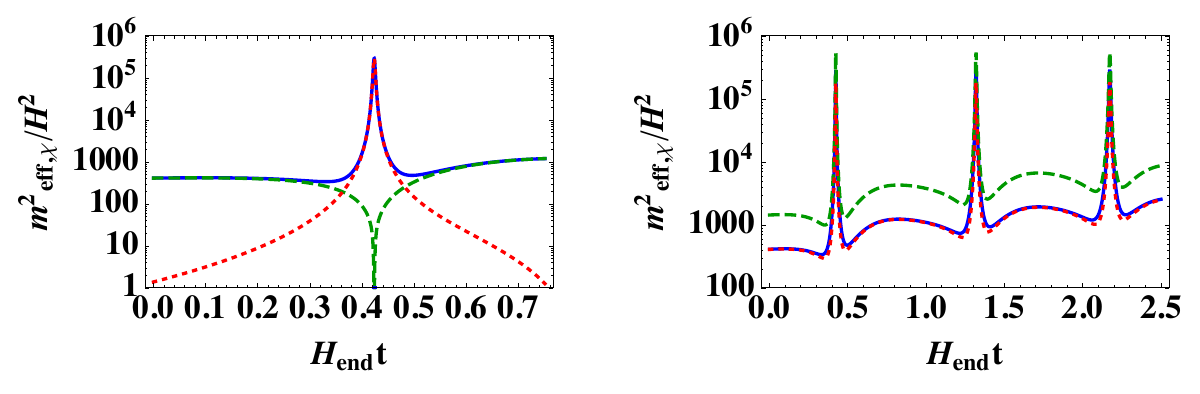}
\caption{ \small \baselineskip 11pt
{\it Left:} Contributions to the effective mass $m^2_{\text{eff},\chi}$ for $\xi_\phi=100$ and $\xi_\chi = 80$ (Case A). The full mass rescaled by the Hubble parameter is shown in blue, $m^2_{1,\chi}/H^2$ in dashed green and $m^2_{2,\chi}/H^2$ in dotted red. The contribution of the spacetime Ricci scalar is suppressed and hence is not visible in the graph. {\it Right:} The effective mass $m^2_{\text{eff},\chi}$ for $\xi_\phi=100$ and $\varepsilon = 0.2, -0.5, 0.5$ (blue, dashed green, dotted red, respectively).}
 \label{fig:effmass}
\end{figure}

The two dominant terms in the effective mass $m_{{\rm eff},\chi}^2$ arise from the potential and the field-space curvature, $m_{1,\chi}^2\equiv  {\cal G}^{\chi\chi}{\cal D}_\chi \partial_\chi V $ and $m_{2,\chi}^2\equiv  -{\cal R}^\chi_{~\phi\phi\chi}  \dot\varphi^2
$ respectively, as they were defined in Ref.~\cite{DeCross:2015uza}. We can approximate the field-space Riemann term by the Lorentzian function of Eq.~\eqref{eq:lorentzian} and the potential term by the simplified form
\beq
m_{1,\chi}^2 \simeq -\tilde \Lambda_\phi \lambda_\phi M_{\rm pl}^2{\varphi^2\over M_{\rm pl}^2+\xi\varphi^2} \, ,
\eeq
 where the denonimator can be set to $M_{\rm pl}^2$, because we are interested in the behavior close to the inflaton zero-crossing.
This allows us to compute simple analytic expressions for the adiabaticity parameter, which depend on the potential parameters and the values of the functions $\varphi(t)$, $\dot\varphi(t)$, $\ddot \varphi(t)$. Fig.~\ref{fig:Lorentzian} shows the result of using the full analytic expression and the numerically computed background evolution $\varphi(t)$. If we only consider the Riemann term, the adiabaticity parameter vanishes for $\varphi(t)=0$, but is otherwise almost constant, with a value of ${\cal A}_{\chi}=2/\sqrt{1-\varepsilon}$. This analytic result can be easily computed using the approximate analytic forms for $m_{1,\chi}^2$ and $m_{2,\chi}^2$. We see that this value is only $10\%$ larger than the numerically computed maximum amplitude of $|{\cal A}_\chi|$, thus it can be used as a quick estimate of the (non-)adiabatic behavior of $\chi$ fluctuations. The left panel of Fig. \ref{fig:effmass} shows the effective mass $m^2_{\text{eff},\chi}$ divided by the Hubble scale for the benchmark Case A with $\xi_\phi = 100$.

\begin{figure}[t!]
\centering
\includegraphics[width=0.95\textwidth]{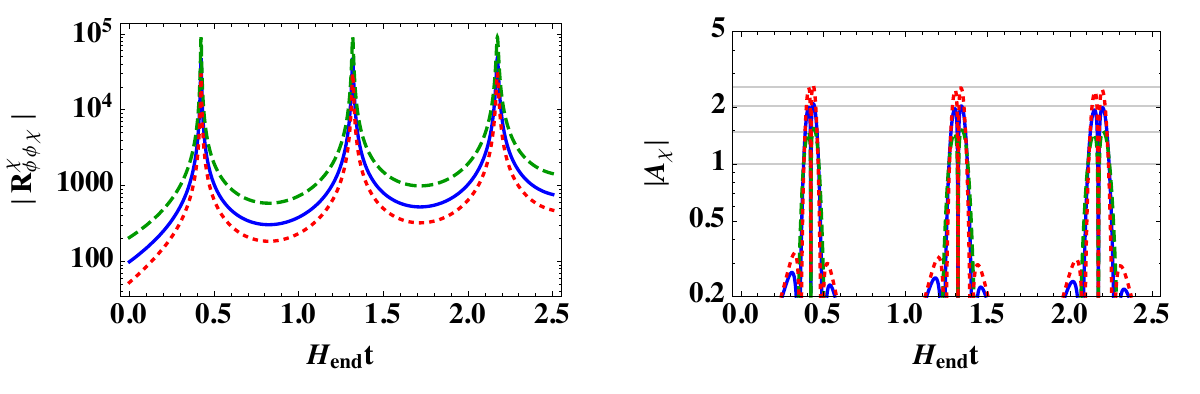}
\caption{ \small \baselineskip 11pt
{\it Left:} The magnitude of the Riemann spike for $\xi_\phi=100$ and $\varepsilon = 0.2, -0.5, 0.5$ (blue, dashed green, dotted red, respectively). 
{\it Right:} The adiabaticity parameter for the same parameters and color-coding.
}
 \label{fig:nadspike}
\end{figure}

Fig.~\ref{fig:nadspike} shows both the Riemann term of the effective mass as well as the (numerically computed) adiabaticity parameter for three values of $\xi_\chi$, corresponding to three different values of the ellipticity $\varepsilon=\pm 0.5$ and $\varepsilon=0.2$. We see that our analytic estimates are in line with the numerical result, and, more importantly, we see that decreasing the value of the ellipticity $\varepsilon$ increases the height of the Riemann spike but reduces the maximum value of the adiabaticity parameter. The corresponding values of the effective mass $m^2_{\text{eff},\chi}$ are shown in the right panel of Fig. \ref{fig:effmass}.

\section{Comparing Lattice Results with Linearized Analysis: Wavenumber Binning Effects}
\label{sec:AppendixCLatticeCompare}

In order to present the results of our lattice simulations, we collect all independent wavenumbers and bin them in intervals of $\Delta k = 2 \pi /L$, where $L$ is the comoving size of the lattice. (A related discussion of sampling effects in lattice simulations of field theory systems related to preheating can be found in Ref.~\cite{Adshead:2016iae}.)
For a three-dimensional lattice of (comoving) size $L$, the corresponding wavenumbers are
\beq
\vec k = {2\pi\over L} (\pm n,\pm m,\pm l) \, ,
\eeq
where $n,m,l$ are positive integers bounded from above by the total number of grid points in each direction, which is equal to $L/\Delta x$, where $\Delta x$ is the (comoving) grid spacing. The number $N$ of independent modes in each bin $k$ scales as $k\sim N^2$ for low $k$, until the magnitude of the wavenumber reaches $k = k_{\rm max}/2$ at which point the number of modes in each bin begins to decrease (related to the Nyquist frequency). This scaling is simply the density of states in a spherical shell in three dimensions.
By choosing the amplitude and phases of the initial conditions of each mode from Gaussian distributions, the resulting power spectra exhibit a variation of $1/\sqrt{N}$. In order to reduce the statistical variance in the final spectra, we bin neighboring wavenumbers by adding squared amplitudes of the various wavenumbers that fit in each bin. Each bin is labeled by the lowest integer wavenumber it contains, so that the $k=10 \, H_{\rm end}$ bin actually contains wavenumbers $10\lesssim k /H_{\rm end} \lesssim 12$.  
As a simple way to properly compare with the lattice results, we compute the following weighted average for the mode $k=10 H_{\rm end}$ 
\beq
|Q_{{k=10H_{\rm end}}}^{\rm lin.}|^2 = {1\over 10^2+11^2+12^2} \left (
10^2 |Q_{k=10H_{\rm end}}|^2  + 11^2 |Q_{k=11H_{\rm end}}|^2 +12^2 |Q_{k=12H_{\rm end}}|^2 
\right ) \, ,
\label{eq:weightaver}
\eeq
with obvious generalization for any wavenumber $k$.
\begin{figure}[t!]
\centering
\includegraphics[width=0.95\textwidth]{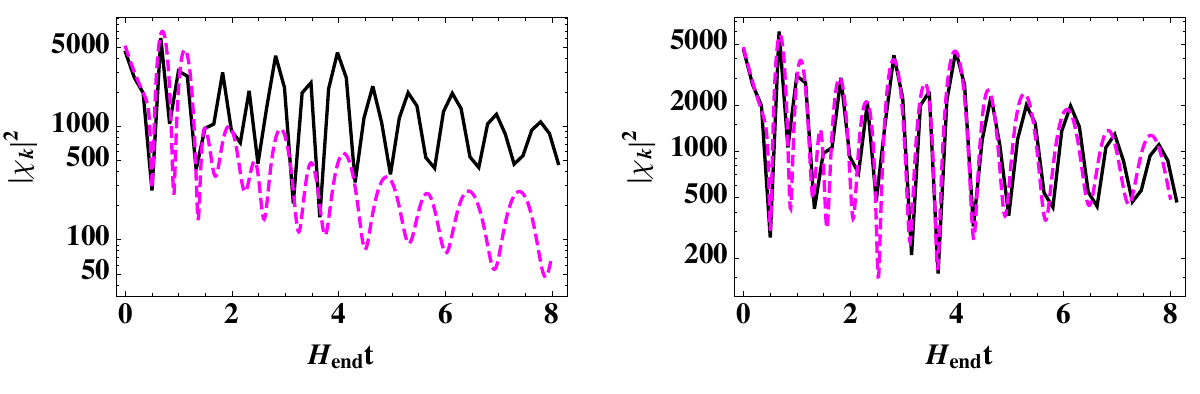}
\caption{\small \baselineskip 11pt
 Comparison of the binned lattice results (black) to the linearized fluctuation analysis (dashed magenta) for the case of positive ellipticity with $\xi_\chi=0.5\xi_\phi$ for $\xi_\phi=10$ and $k=10 \, H_{\rm end}$. On the left, the linear result is shown only using the mode $k = 10 \, H_{\rm end}$, whereas on the right, the linear result is computed using the weighted average of binned modes according to Eq.~(\ref{eq:weightaver}).
}
 \label{fig:AppB}
\end{figure}
We see in Fig.~\ref{fig:AppB} that by using the weighted average procedure of Eq.~\eqref{eq:weightaver} we find excellent agreement between results of the lattice simulations and a linearized analysis (to first order in fluctuations $\delta \phi^I$), as long as the fluctuations have not entered the nonlinear regime. Note that for lower values of $k$, fewer wavenumbers are included in each bin, hence it is more likely to observe large statistical variations in the lattice results.

\section{Excited Wavenumbers and Numerical Convergence Tests}
\label{sec:AppendixDNumericalConvergence}

In this Appendix we consider various requirements on lattice size and other simulation parameters. In order for a lattice simulation to capture relevant dynamics during preheating, the lattice size should be comparable to the Hubble horizon at the end of inflation, hence $k_{\rm min} \lesssim H_{\rm end}$, where $k_{\rm min}$ is the smallest comoving wavenumber in our spectrum (corresponding to the longest wavelength). Meanwhile, for typical values of the couplings, one can show that the modes with comoving wavenumber up to $k_{\rm max} \sim \sqrt{\lambda_\phi} \, M_{\rm pl}$ are typically subject to parametric amplification in this family of models \cite{DeCross:2016fdz,DeCross:2016cbs,Ema:2016dny,Sfakianakis:2018lzf}. During inflation in these models, moreover, the Hubble scale behaves as $H^2 \simeq \lambda_\phi \, M_{\rm pl}^2 / (12 \xi_\phi^2)$, so we have
\beq
{k_{\rm max}\over H_{\rm end} } ={\cal O}(1) \, \xi_\phi \, .
\eeq
In order for a lattice simulation to accurately capture all relevant dynamics of preheating in this family of models, we therefore require ${\cal O} (\xi_\phi)$ grid points in each spatial direction. Thus simulations with $\xi_\phi > 100$ become increasingly expensive.

A further consideration concerns the possible contribution of unphysical ultraviolet (UV) modes to the system's dynamics. Physically, modes that are not excited above the (local adiabatic) vacuum state should not affect the dynamics of the system. Yet classical simulations are based on seeding all modes within the lattice, with initial amplitudes set by the Bunch-Davies (or WKB) values, and then allowing all of the modes to evolve and interact with each other. This can lead to numerical artifacts arising from unphysical contributions from vacuum UV modes. In order to eliminate such contamination in a controlled way, we introduce an initial suppression of the power in small-wavelength modes by multiplying the initial power spectrum by a window function in momentum space of the form
\beq
F(k) = \frac{1}{2} \left[ 1 - {\rm tanh} \left( s \left( k - k_{\rm UV} \right) \right) \right] \, ,
\eeq
where $s$ controls the sharpness of the transition between suppressed and unsuppressed regions of the initial power spectrum, and $k_{\rm UV}$ is the (comoving) threshold wavenumber above which initial power is suppressed. Throughout the main body of this paper, we used $k_{\rm UV} = 50 \, H_{\rm end}$.

\begin{figure}[t!]
\centering
\includegraphics[width=0.95\textwidth]{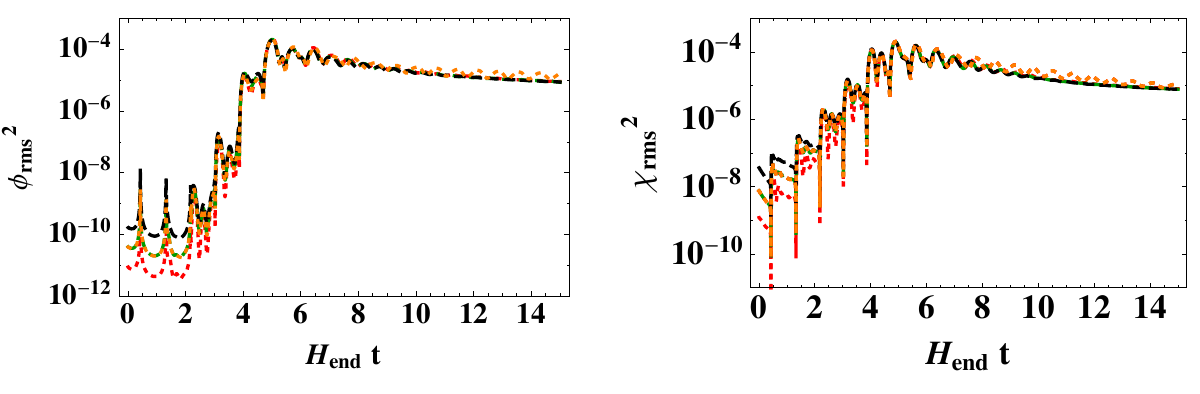}
\caption{\small \baselineskip 11pt 
The variance of the $\phi$ field (left) and the $\chi$ field (right) for the benchmark Case A with $\xi_\phi=100$ and various values of the initial UV suppression scale $k_{\rm UV} /H_{\rm end} = 25,50,100$ (dotted red, dashed green, and dashed black). The late-time behavior of each field is independent of our choice of $k_{\rm UV}$. The orange-dotted curves show the results of a run with ${\cal N} = 512^3$ grid points (rather than ${\cal N} = 256^3$) and $k_{\rm UV} / H_{\rm end} = 50$, again showing excellent agreement between the various simulations.  }
 \label{fig:varCutoff}
\end{figure}

To test the robustness of our numerical results, we performed additional simulations with $k_{\rm UV} / H_{\rm end} = 25, 50, 100$. For small and intermediate values of the nonminimal couplings ($\xi_\phi = 1, 10$), we found that the late-time results were entirely independent of $k_{\rm UV}$, as expected, since in these cases both the linearized analysis \cite{DeCross:2015uza,DeCross:2016fdz,DeCross:2016cbs} and our lattice simulations showed significantly weaker parameter amplification than the case with large nonminimal couplings ($\xi_\phi = 100$). For $\xi_\phi = 100$, we found that the early-time behavior of the variances of the $\phi$ and $\chi$ fields depended on the choice of $k_{\rm UV}$, but that the late-time behavior of each field was essentially independent of $k_{\rm UV}$, as shown in Fig.~\ref{fig:varCutoff}.

We can analyze the effect of varying $k_{\rm UV}$ for $\xi_\phi = 100$ in more detail. For the $\delta \phi$ fluctuations, since no significant parametric resonance is present at early times, increasing $k_{\rm UV}$ simply increases the number of modes that contribute to the initial value of $\langle \phi^2 \rangle$. Since these higher-$k$ modes remain in the vacuum state, their contribution would be removed via renormalization in a fully quantum-mechanical calculation, which is not possible in a (classical) lattice simulation. Nonetheless, we see that the impact of varying $k_{\rm UV}$ vanishes after the first two oscillations of the inflaton condensate (that is, by the third zero-crossing of $\langle \phi \rangle$), after which backreaction and mode-mode interactions dominate the dynamics. For the $\delta \chi$ fluctuations, we find some amplification of modes with large $k$ at early times, and hence some dependence of the early-time field variance on $k_{\rm UV}$. However, just as for the $\phi$ field, by the second oscillation of the inflaton condensate the variance of $\chi$ becomes essentially independent of $k_{\rm UV}$.

In sum, at late times, when nonlinear interactions dominate, both the $\phi$ and $\chi$ spectra are independent of the choice of $k_{\rm UV}$. In particular, the nonlinear dynamics generate a strong cascade of power toward higher-$k$ modes, yielding a smooth final spectrum for each field, which exhibits no memory of the initial window function or the value of $k_{\rm UV}$ that had been used.

Likewise, we find very little dependence of our numerical results as we vary the number of grid points ${\cal N}$, as shown in Fig.~\ref{fig:varCutoff}. If we keep the UV suppression scale fixed at $k_{\rm UV} = 50 \, H_{\rm end}$ and change from ${\cal N} = 256^3$ to $512^3$, the behavior of the system is essentially indistinguishable for times through $H_{\rm end} \, t \simeq 6$, which is when preheating completes for this set of couplings. After that time, the results of the two simulations track each other well, though the simulation with ${\cal N} = 512^3$ exhibits more visible oscillations in the variances of each field. This is related to the strong cascade of power toward higher-$k$ modes: the increased number of grid points allows us to capture a larger part of this UV cascade. However, this very slight difference at late times has no effect on the physical quantities of interest, such as duration of reheating $N_{\rm reh}$ or the effective equation of state. For smaller values of $\xi_\phi$, for which the UV cascade is weaker, the late-time results show even less variation with ${\cal N}$.

Finally, it is worth noting that the problem studied here is well within the the regime in which a semiclassical approximation is valid.  In all of the cases in which these nonperturbative effects occur, the system we study is within the region of validity of a {\it classical}, real-time treatment (e.g., exhibiting large occupation numbers), as originally discussed in Refs.~\cite{Khlebnikov:1996mc,Prokopec:1996rr}. A detailed description of contemporary lattice methods can be found in Ref.~\cite{Figueroa:2020rrl}.


%

\end{document}